\documentclass[preprint2]{aastex}

\shorttitle{On the prominence of spiral galaxy bulges}
\shortauthors{A.W.GRAHAM}

\begin{document}

\input{psfig}

\title{AN INVESTIGATION INTO THE PROMINENCE OF SPIRAL GALAXY BULGES}
\author{Alister W. Graham\altaffilmark{1}}
\affil{Instituto de Astrof\'{i}sica de Canarias, La Laguna, 
E-38200, Tenerife, Spain}
\email{agraham@ll.iac.es}
%\email{agraham@ll.iac.es, mpm@ll.iac.es} 

\altaffiltext{1}{Isaac Newton Group, La Palma, Spain}

\begin{abstract}
From a diameter-limited sample of 86 `face-on' spiral galaxies, 
the bulge-to-disk size and luminosity ratios, and other quantitative 
measurements for the prominence of the bulge are derived.  The bulge 
and disk parameters have been estimated using a seeing convolved 
S\'{e}rsic $r^{1/n}$ bulge and a seeing convolved exponential disk 
which were fitted to the optical ($B, R$, and $I$) and near-infrared ($K$) 
galaxy light profiles.  In general, early-type spiral galaxy bulges 
have S\'{e}rsic values of $n$$>$1, and late-type spiral galaxy bulges 
have values of $n$$<$1.  Use of the exponential ($n$$=$1) bulge model 
is shown to restrict the range of $r_e/h$ and $B/D$ values by more than 
a factor of 2.  Application of the $r^{1/n}$ bulge models
results in a larger mean $r_e/h$ ratio for the 
early-type spiral galaxies than the late-type spiral galaxies.  Although, 
this result is shown not to be statistically significant.  The mean $B/D$ 
luminosity ratio is, however, significantly larger ($>$3$\sigma$) for the 
early-type spirals than the late-type spirals.  

Two new parameters are introduced to measure the prominence of the bulge.
The first is the difference between the central surface brightness of the 
galaxy and the surface brightness where the bulge and disk contribute
equally.  The other test uses the radius where the contribution from the 
disk and bulge light is equal -- normalised for the effect of intrinsically 
different galaxy sizes.  Both of these parameters reveale that 
the early-type spiral galaxies `appear' to have significantly 
($>$2$\sigma$ in all passbands) bigger and brighter bulges than 
late-type spiral galaxies. 
This apparent contradiction with the $r_e/h$ values can be explained 
with an iceberg-like scenario, in which the bulges in late-type spiral
galaxies are relatively submerged in their disk.  This can be
achieved by varying the relative bulge/disk stellar density while 
maintaining the same effective bulge-to-disk size ratio.

The $B/D$ luminosity ratio and the concentration index $C_{31}$ are, 
in agreement with past studies, positively correlated and decrease as 
one moves along the spiral Hubble sequence towards later galaxy types.
Although for galaxies with large extended bulges, the concentration index
no longer traces the $B/D$ luminosity ratio in a one-to-one fashion. 
A strong (Spearman's $r_s$=0.80) and highly significant positive 
correlation exists between the shape, $n$, of the bulge light profile 
and the bulge-to-disk luminosity ratio.  

The absolute bulge magnitude -- $\log n$ diagram is used as a diagnostic
tool for comparative studies with dwarf elliptical and ordinary elliptical
galaxies.  At least in the $B$-band, these objects occupy distinctly 
different regions of this parameter space.  While the dwarf ellipticals 
appear to be the faint extension to the brighter elliptical galaxies, 
the bulges of spiral galaxies are not; for a given luminosity they have 
a noticeably smaller shape parameter and hence a more dramatically declining
stellar density profile at larger radii.

\end{abstract}


%XXX
\notetoeditor{
1. I would like Figure~\ref{seeing} to run the full width of a page
rather than just one column.  
2. The Appendix figure runs over 7 pages. 
3. Thanks}

\section{Introduction}

Hubble (1926, 1936) used three criteria to classify spiral galaxies 
in what has become known as the Hubble sequence.  Based on the structural 
forms of photographic images, his first criteria was the ``relative size
of the unresolved nuclear region''.   The other criteria 
were the degree of resolution in the arms, and 
the extent to which the spiral arms are unwound. 
Going from early-type (Sa) to late-type spirals (Sc in Hubble's 
classification), Hubble wrote that 
``{\it the arms appear to build up at the expense of the nuclear regions 
and unwind as they grow; in the end, the arms are wide open 
[highly resolved] and the nuclei inconspicuous}''.\footnote{Hubble 
also used an additional classification, which was the presence of a bar, 
giving rise to a parallel sequence of spiral galaxies that followed the
above criteria and led to the second prong in the Hubble tuning-fork
diagram.}
% (XXX - check this in his 1936 book)

\begin{figure}
\centerline{\psfig{figure=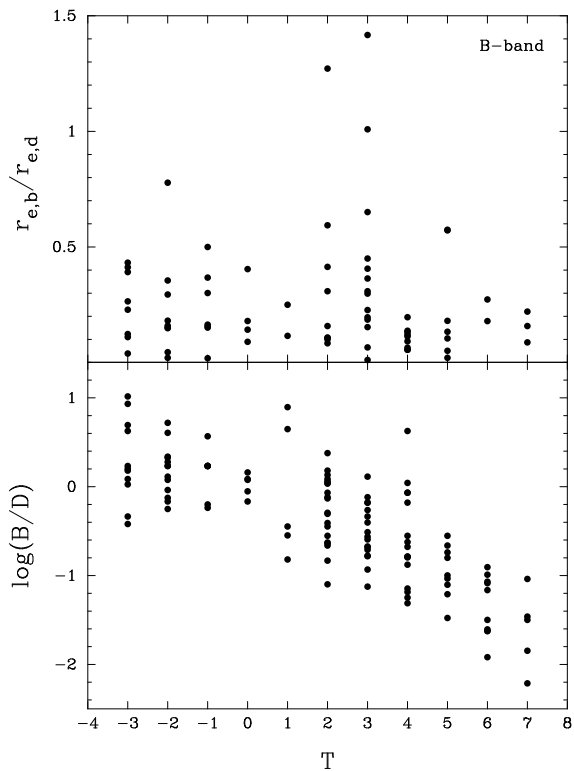,width=7.5cm,angle=00}}
\caption{This figure shows the bulge-to-disk effective radius ratio
$r_{e,b}/r_{e,d}$ (upper panel) and the logarithm of the bulge-to-disk
luminosity ratio $B/D$ (lower panel) from the data of Simien \& de
Vaucouleurs (1986).}
\label{amazing}
\end{figure}

With the above order of criteria reversed, and basing spiral galaxy 
classification primarily on
the characteristics of the arms, Sandage (1961) notes in the `Hubble Atlas
of Galaxies' that Sa type galaxies can exist with both large and small 
bulges, however, a general correlation still exists between the relative 
size of the bulge and the criteria of the arms (or morphological type).  
For a review of the Hubble classification scheme, see van den Bergh (1997).
Simien \& de Vaucouleurs (1986) were bold enough to try placing this 
trend of decreasing bulge-to-galaxy luminosity ratio with increasing galaxy 
type on a quantitative basis -- fitting a cubic to the observed relation.  
In order to do this, they compiled a sample of 98 galaxies (from six different 
sources) with galaxy type -3$\leq$$T$$\leq$7 -- 64 with type Sa or later. 
They modelled the bulges with an $r^{1/4}$ law and the disks with an 
exponential profile.  Commenting on the standard deviation of 1.14 mag 
between multiple observations of the same galaxies\footnote{The standard 
deviation of 1.14 mag from Simien \& de Vaucouleurs (1986) refers to the 
measured bulge magnitude minus the total galaxy magnitude between similar 
galaxies observed by different authors.  This is the quantity used in their 
figure 2 and 3.}, they wrote that ``Clearly, there is room for 
improvement''.  
The bulge-to-disk size and luminosity ratios from their paper 
is presented here in Figure~\ref{amazing}.\footnote{The $\Delta m_I$ data 
from table 4 of Simien \& de Vaucouleurs (1986) has been used to 
compute the $B/D$ luminosity ratios shown here in Figure~\ref{amazing}.}

%\placefigure{amazing}

More recently, using an exponential (Patterson 1940; Freeman 1970) bulge
profile rather than a de Vaucouleurs (1948, 1959) $r^{1/4}$ bulge profile, 
de Jong (1996b) and Courteau, de Jong, \& Broeils (1996) claimed that the 
bulge-to-disk scale-length (or size) ratio was in fact independent of 
Hubble type -- a result consistent with the data of Simien \& de 
Vaucouleurs (1986).  If the bulge-to-disk size ratio is statistically 
shown to be independent of morphological type, this would imply that the 
variation in the bulge-to-disk luminosity ratio that characterises the 
Hubble sequence for spirals is either due to changes in the relative surface 
brightness of the bulge and disk, and/or changes in the bulge profile shape. 

However, before one draws any foregone conclusions, D'Onofrio, Capaccioli, 
\& Caon (1994) have already questioned the applicability of a fitting-function 
which cannot account for possible variations in the curvature of a
bulge's light profile, and it has since been shown that spiral galaxy 
bulges are in fact not universally described with either an exponential profile 
or an $r^{1/4}$ law (Andredakis, Peletier, \& Balcells 1995; Moriondo, Giovanardi, 
\& Hunt 1998; Khosroshahi, Wadadekar, \& Kembhavi 2000).  Rather, a continuous 
range of bulge light profile shapes are now known to exist. 
These have been well described by the S\'{e}rsic (1968) 
$r^{1/n}$ profile\footnote{The S\'{e}rsic model reduces to the $r^{1/4}$ law when 
$n$$=$4, and reduces to an exponential profile when $n$$=$1.}, and, moreover, 
it has been shown that early-type spiral galaxy bulges have larger shape 
parameters $n$ than late-type spiral galaxy bulges.  

In fact, in the profile fitting performed by de Jong (1996a), he found that 
40\% of the spiral galaxy bulges where better modelled with an $n$$=$2 
or $n$$=$4 model than an exponential model.  He statistically showed that, 
on average, the Sa and Sb galaxies are better modelled with an $r^{1/2}$ 
bulge light profile than an exponential profile and that the Sbc-Scd galaxies 
are fit equally as well with the $n$=2 bulge model as the $n$=1 bulge model 
(see Figure 4 of de Jong 1996a).  
Furthermore, from the 30 early-type disk galaxies modelled with a S\'{e}rsic
bulge by Andredakis et al.\ (1995), only two had shape parameters $n$ consistent 
with a value 1 or less. 
Only the late-type ($\geq$Sd) spiral galaxies in the sample of de Jong
were better modelled, on average, with an $n$$=$1 profile and possibly
 even these have true 
bulge profile shapes significantly different from a pure exponential.   
Therefore, any investigation into the global bulge-to-disk 
properties of spiral galaxies that ignores this structural trend will be biased
to some degree.   Moriondo, Giovanardi, \& Hunt (1998) also drew attention to
this with a sample of 14 galaxies, revealing how forcing an exponential 
bulge profile can restrict the full range of galaxy parameters which exist.  

Using the model parameters from the `best-fitting' $r^{1/4}$, 
$r^{1/2}$, or 
exponential profile from each galaxy in de Jong's sample, (as determined using 
the $\chi ^2$ values from de Jong 1996a), Graham \& Prieto (1999a,b) 
reinvestigated the claim for a scale-free Hubble sequence.  
Using the $K$-band data set of de Jong (1996a), Graham \& Prieto (1999a) showed 
that the average bulge-to-disk scale-length ratio obtained using 
the exponential bulge model is actually {\it smaller} (at the 98\% confidence level)
for the early-type spiral galaxies than the late-type spiral galaxies.  In fact, 
for all passbands used, use of the exponential bulge profile resulted in an 
average $r_e/h$ ratio that was significantly ($\sim$2-3$\sigma$) smaller for 
the early-type spiral galaxies than that obtained when using the 
best-fitting profile models.  Consequently, 
failing to account for the different structural profiles, which are
dependent on morphological type, seriously affects ones
ability to draw any subsequent conclusions about trends between
structural properties and galaxy type.

This paper presents a further, more detailed, analysis into the claim that the 
bulge-to-disk size ratio is independent of morphological type, and goes on to 
explore the bulge-to-disk luminosity ratio. 
Rather than simply using the $n$=1, 2, or 4 bulge models, the optical 
($B$, $R$, $I$) and near-infrared ($K$) light profiles from de Jong (1996a) 
are re-modelled using a seeing convolved S\'{e}rsic $r^{1/n}$ bulge and a seeing 
convolved exponential disk -- as described in Section 2.  The galaxy 
sample and best-fitting model parameters are presented in Section 3. 
Section 4 explores both the bulge-to-disk size and luminosity ratio 
as a function of morphological type.   
A preliminary analysis was briefly reported in Graham \& Prieto (2000a,b). 
In Section 5 other quantitative parameters of bulge strength, 
such as the concentration index, and two new parameters which reflect the
visual appearance of the bulge are explored.  A summary and conclusions are presented
in Section~\ref{SecCon}.

\section{S\'{e}rsic light profiles}

\subsection{The effects of seeing on $r^{1/n}$ light profiles}

\begin{figure}
\centerline{\psfig{figure=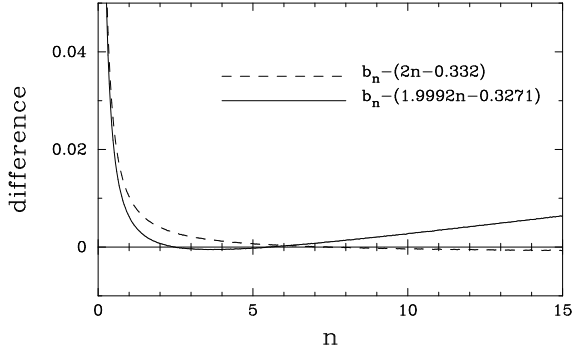,width=7.5cm,angle=-90}}
\caption{This figure shows the difference between the exact value for $b_n$
from the S\'{e}rsic $r^{1/n}$ model (equation~\ref{eq_ser}), such that
$\Gamma (2n)$$=$2$\gamma (2n,b_{n})$, and popular approximations used for
elliptical galaxies where $n$$\sim$4.}
\label{butt-fig}
\end{figure}

The S\'{e}rsic (1968) $r^{1/n}$ radial intensity profile can be written as 
\begin{equation}
I(r)=I_{e}\exp \left[ -b_n \left\{ \left( \frac{r}{r_e} \right) ^{1/n} -1 \right\} \right] 
\label{eq_ser}
\end{equation}
where $I_{e}$ is the intensity at the effective radius, $r_e$, which encloses  
50\% of the light.  The term $b_n$ is a function of the shape parameter $n$, 
such that $\Gamma (2n)$=2$\gamma (2n,b_{n})$, where $\Gamma $ is the gamma function 
and $\gamma $ is the incomplete gamma function.  As given by Capaccioli (1989), this
can be well approximated by $b_{n}$$=$$1.9992n-0.3271$ for 1$<$$n$$<$10.
This approximation becomes worse for values of $n$$<$1 -- as can be seen in
Figure~\ref{butt-fig}.  Given that many of the spiral galaxy bulges in this sample 
turn out to have values of $n$$<$1, the exact expression for $b_n$ has been used 
instead of the above approximation.  To simplify the appearance of the 
following equations, the subscript $n$ will be dropped from the term $b_n$.

%\placefigure{butt-fig}

Corrections for the effects of seeing have been made using the prescription 
given in Pritchet \& Kline (1981).  For any intrinsically radially symmetric, 
intensity distribution $I(r)$, the seeing convolved profile, $I_c(r)$, is such that 
\begin{equation}
I_{c}(r)=\sigma^{-2}e^{-r^2/2\sigma ^2}\int ^{\infty}_{0} I(x)I_0(xr/\sigma^2)
e^{-x^2/2\sigma ^2}xdx,
\label{EqConv}
\end{equation}
where $\sigma $ is the dispersion of the Gaussian PSF (=FWHM/2.3548), and
$I_0$ is the zero-order modified Bessel function of the first kind (e.g.\
Press et al.\ 1986).  This approach at correcting the bulge (and 
disk) profile for seeing was adopted by Andredakis et al.\ (1995) 
and later de Jong (1996b).\footnote{n.b.\ de Jong (1996b) omitted the 
negative sign from the final exponent.}
% (For non-spherical 2D images, such as elliptical galaxies, Trujillo et al.\ (2000) 
% provide a fuller prescription to derive the seeing corrected S\'{e}rsic model 
% parameters.)
In this paper, the convolution of equation~\ref{EqConv} is applied to both S\'{e}rsic
bulge models with free shape parameters (i.e.\ not fixed to integer values) 
and also to the exponential disk profiles. 
These convolved models are then simultaneously fit to the light profile 
data using a standard non-linear least-squares algorithm, which is iterated until 
convergence on the optimal solution giving the smallest $\chi ^2$ value. 

\begin{figure}
\centerline{\psfig{figure=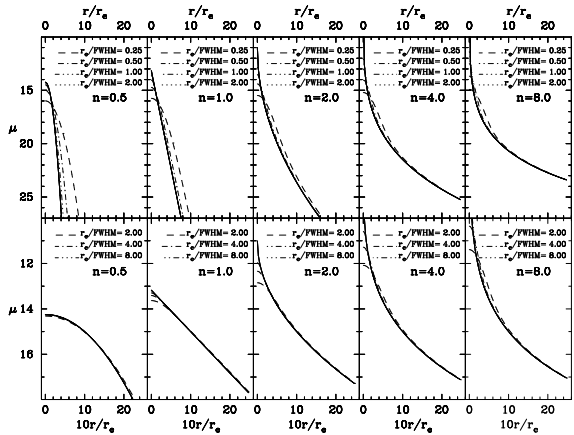,width=7.5cm,angle=-90}}
\caption{The effects of seeing on the S\'{e}rsic $r^{1/n}$ profile.
The solid lines show a series of $r^{1/n}$ models with $n$=0.5, 1.0, 2.0, 4.0,
and 8.0.  The broken lines show the convolution of each model with a
Gaussian PSF having a range of ratios between the effective half-light radius
of the $r^{1/n}$ model and the FWHM of the Gaussian PSF.  The lower panel
shows an enlarged view of the inner two effective radii when the effects of
seeing are more subtle.}
\label{seeing}
\end{figure}

Figure~\ref{seeing} shows the effects of seeing on various S\'{e}rsic profiles with
different shape parameters. 
What is important is the ratio between the dispersion (or FWHM) of the PSF
and the effective half-light radii ($r_e$) of the S\'{e}rsic model. 
Figure~\ref{seeing} has been designed to highlight the most dramatic cases of 
how seeing can affect the S\'{e}rsic luminosity profile.  
When $n$ is small (e.g.\ 0.5) and there is a small $r_e$/FWHM 
ratio, the original light profile is somewhat akin to a point source
and the seeing convolved light profile therefore looks like the PSF. 
Similarly, when  $r_e$/FWHM is small (e.g.\ $<$2) and the intrinsic light 
profile falls away quickly with radius, away from the center the seeing convolved 
profile can be substantially brighter than original profile as the `blurred' 
light from the brighter inner radii dominates at larger radii.  While this may
look quite severe in some cases, this effect would be substantially more dramatic if 
a Moffat (1969) function with say $\beta $$=$2.5, or any other PSF which has 
higher wings than a Gaussian (e.g.\ Saglia et al.\ 1993, their Figure 1) was 
used.  For the galaxy sample used, all of the derived bulge values for $r_e$ are 
greater than 1$\arcsec$, and the ratio $r_e/$FWHM is greater than 1 for all but 
three galaxies, and for these galaxies the ratio is only just less than 1.

%\placefigure{seeing}

\subsection{The S\'{e}rsic bulge to exponential disk luminosity ratio}

The ratio of the bulge-to-disk luminosity $(B/D)$ is an important quantity for
studies into spiral galaxies.  The trend of decreasing $B/D$ luminosity ratio 
from early- to late-type spirals has since been quantified by, amongst others, 
Kent (1985) and Simien \& de Vaucouleurs (1986) and is regarded as one of the 
prime characteristic of the Hubble sequence for spiral galaxies.  

The total luminosity described by a S\'{e}rsic $r^{1/n}$ light profile is given 
in Graham et al.\ (1996) as 
\begin{equation}
L_{\rm tot} = \int _{0}^{\infty} I(r)2\pi rdr = 
\frac{n2\pi r_{\rm e}^{2}I_{\rm e}\,e^{b}}{b^{2n}}\Gamma (2n),
\end{equation}
where $r_{\rm e}$ is in arcseconds. 
When $n$=4, one obtains the familiar de Vaucouleurs expression
$L_{\rm tot}$=7.215$\pi I_e r_e^2$ and when $n$=1 one obtains 
$L_{\rm tot}$=3.803$\pi I_e r_e^2$. 
The total apparent magnitude $m_{\rm tot}$ is simply -2.5$\log L_{\rm tot}$, and the
total absolute magnitude $M_{\rm tot}$ given by $M$$=$$m-5\log (D) -25$, where $D$
is the distance to the galaxy in Mpc. 

However, for the disk, rather than use $r_e$ and $I_e$, it is common practice
to use the disk scale-length ($h$=$r_e/b^n$) and the central surface 
brightness $I_0$=$I_e e^b$. 
From here on, the use of $r_e$ and $I_e$ will refer only to that of the 
bulge.  The ratio of bulge-to-disk luminosities is then given by the expression 

\begin{equation}
\frac{B}{D}=\frac{n_{\rm b}\Gamma (2n_{\rm b}) e^{b}/b^{2n_{\rm b}}}{n_{\rm d}\Gamma (2n_{\rm d})}
\left( \frac{r_{\rm e}^{2}}{h^2} \right)
\left( \frac{I_{\rm e}}{I_0} \right),
\label{MuEq}
\end{equation}

where the subscripts b and d on the shape parameter $n$ refer to the bulge 
and disk respectively, and the parameter $b$ ($=$$b_n$) refers to that of 
the bulge.  Given that disks are well described by an exponential profile, 
one can substitute $n_{\rm d}$ with the value of 1.  For integer values of $x$, 
$\Gamma (x)$=($x$-1)! and equation~\ref{MuEq} can be further reduced to give

\begin{equation}
\frac{B}{D}=\frac{(2n_{\rm b})! e^{b}}{2.b^{2n_{\rm b}}}
\left( \frac{r_{\rm e}^{2}}{h^2} \right)
\left( \frac{I_{\rm e}}{I_0} \right), 
\end{equation}

where $n_{\rm b}$ refers to the shape parameter of the bulge.
When $n_{\rm b}$=4, the factor in the front reduces to 1/0.28 as given in 
Binney \& Merrifield (1998). 

It is interesting to note that if the Hubble sequence for spiral galaxies 
is shown to be scale-free (i.e.\ independent of the $r_e/h$ ratio), 
then any variations in the $B/D$ luminosity ratio must be due to changes in the 
relative surface brightness of the bulge and disk, and/or the shapes 
of the light profiles of these components. 

\begin{figure}
\centerline{\psfig{figure=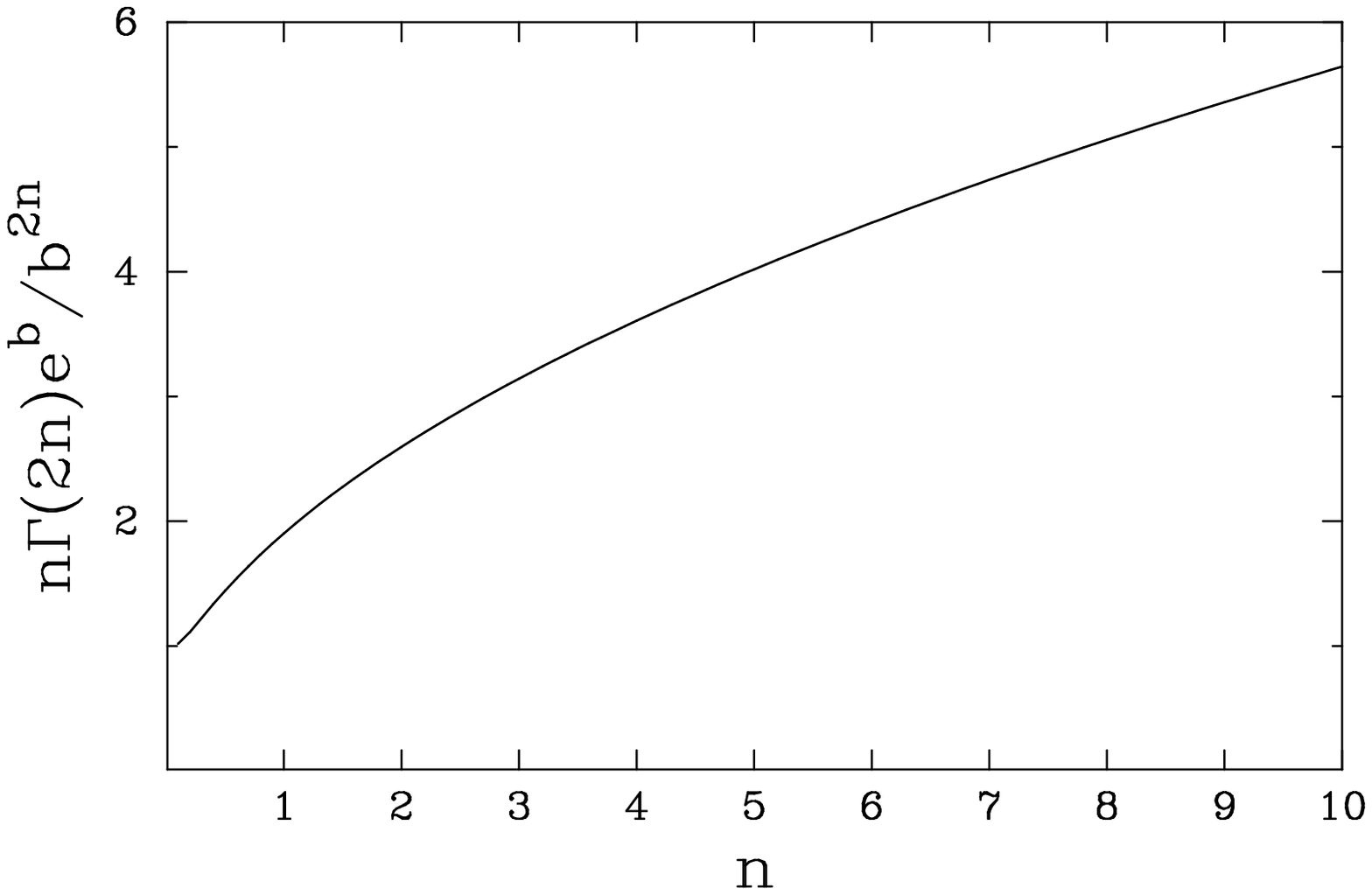,width=7.5cm,angle=-90}}
\caption{This plot reveals the bulge-to-disk luminosity dependence on the bulge
shape parameter $n$, as given in equation~\ref{MuEq} (with $n_{\rm d}$=1).}
\label{Fnfig}
\end{figure}

For the sake of curiosity, to investigate how the first term in 
equation~\ref{MuEq} varies with $n_{\rm b}$, Figure~\ref{Fnfig} 
shows the relationship between 
$(n_{\rm b}\Gamma (2n_{\rm b}) e^{b}/b^{2n_{\rm b}})/n_{\rm d}\Gamma (2n_{\rm d})$
and $n_{\rm b}$, with $n_{\rm d}$$=$1.

%\placefigure{Fnfig}

\section{The Data}

\subsection{The Galaxy Sample}

The one-dimensional light profile data\footnote{The data can be found at
\url{http://cdsweb.u-strasbg.fr/htbin/Cat?J/A+AS/118/557}.}  from the 
statistically complete diameter-limited sample of 86 undisturbed 
low-inclination (face-on) spiral galaxies presented in de Jong \& van der 
Kruit (1994) is reanalyzed here.  
A detailed discussion of the selection criteria, observations, and data 
reduction can be found in their paper.  In this study, the optical passbands 
$B$, $R$, and $I$, and the near-infrared $K$-band data are used. 
The emphasis will be placed on the near-infrared data set
because it is the best tracer of a spiral galaxy's global luminous 
structure -- avoiding the obscuring effects of dust and biasing from a
few percent (by mass) of hot young stars.  The $B$-band is presented
for comparison with both the $K$-band data presented here and previous 
studies of this nature. 

Of the 86 galaxies, there is no K-band data for three (UGC 02125, UGC 02595, 
UGC 12808) and a further five galaxies (UGC 10437, UGC 11708, UGC 12732, 
UGC 12754, UGC 12845) were imaged under non-photometric 
conditions.  However, for a number of tests performed here it does not 
matter if the data is photometrically calibrated or not as scale-lengths and 
bulge-to-disk surface brightness differences are used.  In tests
where this is important, these five galaxies have been excluded. 
Three galaxies could not be reliably modelled: UGC 08279 and UGC 12732 
had no discernable bulge, and the disk of UGC 6028 has two distinct slopes. 
UGC 09024 was also excluded because NED gives it's morphological type
only as `S?', and the Sa galaxy UGC 00089 has been excluded as it's bulge 
parameters are heavily biased by the presence of a strong bar. 
At this stage all the remaining galaxies are included.  Later on, the 
effect of collectively removing all galaxies with a prominent bar is explored. 
Therefore, either 78 or 74 galaxies were used in the following
$K$-band analysis -- depending on whether or not photometric data was required. 
In the $B$-band, this rejection process also left either 78 or 74 galaxies. 
% In the $B$-band UGC 2595 has no data; UGC 00242 and UGC 10437 could 
% not be modelled, nor could UGC 08279 and UGC 12732, and UGC 6028 was 
% rejected.

The morphological types listed by de Jong (1996a) agree well with the year 
2000 database in NED.  For only three galaxies is there a difference. 
de Jong lists UGC 10437 as T=5, NED as T=7; UGC 10445 as T=5, NED as T=6; 
and UGC 09024 as T=8, and NED as S?  The most up-to-date values have been taken, 
and UGC 09024 excluded from the final analysis as it's exact Hubble type is 
not specified in NED. 

% The statistical analysis in this paper will be performed using every galaxy, 
% and then again excluding those galaxies with a bar, and then a third time, 
% excluding those galaxies which required truncation of the
% outer data points (see Section~\ref{model}). 

% Given that this analysis is done using the (Sa,Sab,Sb) 
% type galaxies versus the (Scd,Sd,Sdm,Sm) type galaxies (excluding the 
% intermediate Sbc and Sc type galaxies), only five galaxies will be excluded 
% in the final analysis because they were truncated.  

\subsection{Surface brightness corrections}

The $r_e/h$ ratio is of course independent of the photometric zero-point. 
However, the need for photometric corrections are required in this investigation 
as the $K_{22}$ radius (the radius where the $K$-band surface brightness equals 
22.0 mag arcsec$^{-2}$) is used in an attempt to normalise the size of each galaxy. 
Also, in the $K$-band, galaxy disks are transparent and their surface 
brightness depends on the inclination of the galaxy.  Therefore, if one 
hopes to compare galaxies at different inclinations, one must correct 
the disk surface brightness to some standard, such as the face-on value. 

The standard surface
brightness inclination correction is given by the expression 2.5$C\log(a/b)$, 
where $a/b$ is the ratio of the semi-major over semi-minor axis.
For a transparent galaxy, the more inclined it is, the greater the
line-of-sight through the galaxy and hence the surface brightness
will appear brighter than it would if the galaxy was face on.  This
corrective term is therefore applied to reduce the observed surface brightness.
The value of $C$ is equal to 0 for an optically
thick disk, and equal to 1 for a transparent disk.
For the K-band data from this low-inclination sample of spirals, 
the value of $C$ is set equal to 1 (Graham 2000).  For the $B$-band $C$$=$0 has been used.  
For the $R$- and $I$-bands the $C$ values from Tully \& Verheijen (1997) are used, 
which are 0.52 and 0.61 respectively.  

Corrections for Galactic extinction were made using the extinction data,
presented in NED, from the composite IRAS and COBE/DIRBE dust
extinction maps of Schlegel, Finkbeiner \& Davis (1998).
% These reddening measurements are significantly more 
% accurate than the estimates from Burstein-Heiles (1984) -- which
% were based on galaxy counts and neutral hydrogen emission.  
For the sample of spirals used here, these new reddening estimates are
noticeably larger than the Burstein-Heiles estimates.  The mean
reddening value in the B-band is 0.24 mag with the largest correction 1.33 mag
(c.f.\ an average value of 0.14 mag with the Burstein-Heiles data).
As Schlegel et al.\ (1998) used an $R_V$=3.1 extinction curve,
$A_K$$=$$0.085A_B$ which
translates into an average reddening correction of only 0.02 mag in the K-band
with the maximum K-band correction 0.11 mag.

Two small additional corrections which de Jong (1996b) did not apply are 
applied here.  The first is the $(1+z)^4$ cosmological redshift dimming.
The galaxy sample extends to distances of 8000 km s$^{-1}$, where
the cosmological surface brightness dimming, given by
10$\log(1+z)$, is not insignificant at 0.114 mag.
The distances tabulated by de Jong (with $H_0$$=$100 km s$^{-1}$
Mpc$^{-1}$ and allowing for Virgo infall) were converted
into redshifts that were used to calculate this correction. 

The second correction is the k-correction.  Due to the stretching
of wavelength with redshift, fixed passbands at the
telescope sample different intrinsic wavelengths from galaxies
at different redshifts.   Heliocentric velocities were used
to correct for this -- the difference between the redshift observed at
the telescope and the Sun is, of course, not significant for this correction.
The tables in Poggianti (1997) were used for this correction.  No
correction for evolution was applied. 

A Hubble constant of $H_0$$=$75 km s$^{-1}$ Mpc$^{-1}$ has been used in the 
final conversion from arcsec to kpc, and from apparent magnitude to absolute magnitude.

\subsection{The best-fitting model parameters}
\label{model}

One of the early methods for parametrizing the disk involved `marking the disk', 
by eye, over the radial range where it's profile appeared to be linear.  
Not surprisingly, by construction, this approach gave results that `appeared'
to be correct; and initself there is nothing wrong with this.  However, at a basic
level, because spiral galaxy light profiles are the superposition of both a disk and 
a bulge component\footnote{At a more refined level, spiral galaxy light profiles are 
the superposition of several components, such as, in addition to the dominant 
bulge and disk componemts: lenses, rings, and bars, which, from a careful 2D image 
analysis, can also be modelled for some galaxies (Prieto et al.\ 1997, 2000).}, 
the inner part of disk, which may appear to be linear 
with the outer disk profile, can be biased by the presence of light from the outer 
parts of the bulge.  Such problems with the `marking the disk' method are well 
known (Giovanelli et al.\ 1994) and so both components should be fit simultaneously
(Kormendy 1977) as was done by de Jong (1996a).  One must then decide what 
structural form to use for the bulge.  The estimated central surface 
brightness of the disk, obtained when simultaneously fitting the disk with first an 
$r^{1/4}$ bulge and then an exponential bulge can vary by half a magnitude and
sometimes much more (Graham 2000).  The differences to the bulge parameters are 
even more dramatic, and so exactly which bulge profile shape one uses is an 
important consideration.  

Although de Jong (1996a) noted that a range of bulge profile shapes do 
exist, and that this together with the sky backgound uncertainty are the two
most important sources of error, he presented reasons for having not modelled 
these structural differences.  Firstly, for most of the galaxies, his code for 
fitting an $r^{1/n}$ bulge model converged on physically unacceptable negative 
values for the model parameters.  
This problem was solved here by the inclusion of boundaries to the 
parameter space which was searched -- 
as done by Schombert \& Bothun (1987) who recognised the need for physical 
limits to what is a purely numerical method.  Secondly, de Jong commented 
that for many 
of the late-type spiral galaxies the bulge light dominates over the disk light 
for only a few data points which makes it hard to accurately limit the shape
of the bulge.  This is true for galaxies with small bulges and so error estimates 
are useful to gauge this problem.  However, it is noted that in fitting the 
bulge and disk simultaneously, the fitting of
the bulge is not solely restricted to those few inner points, but is additionally 
constrained by data points further out -- even though the disk dominates in this region. 
% Despite this, for ``consistency reasons'' de Jong chose to fit an exponential bulge 
% to every galaxy, whether or not the errors on the bulge profile shape could be well 
% constrained.  The extent of possible systematic biases which this may cause are 
% explored in this paper.  
%However, even if one feels that they {\it are} unable to constrain the shape of the 
%bulge for some galaxies, in which case the errors on the parameters would be good
%to know, one would then be faced with the question of what profile shape to assume. 

For each galaxy, the best-fitting seeing-convolved S\'{e}rsic bulge model and 
exponential disk model were derived three times.  The difference was that in 
the second and third derivation, the uncertainty in the sky-background 
level was respectively added and subtracted from the light profile data. 
This was also the approach used by de Jong (1996a) to estimate the errors
on the model parameters.  For a handful of galaxies, subtraction of the 
sky-background error resulted in a negative intensity for some of the outer 
data points in the light profile.  When this happened, these points were removed 
and the fitting proceeded as normal.  Similarly, the error on the $B/D$ luminosity 
ratio represents the $B/D$ luminosity ratio obtained when the uncertainty 
to the sky-background level was added and subtracted from the light profile. 

One source of error which is not dealt with explicitly in this analysis of the 
1D light profiles is the presence of third components -- such as a bar.  For 23 
of the 86 galaxies modelled by de Jong, the introduction of bar improved his 2D 
fits at some level.  The statistical analysis which follows is however performed 
including and excluding these galaxies.  Sometimes this had an effect, and other 
times the reduced number statistics simply weakened the significance of a 
result. 
Another issue pertaining to the model parameters is that of coupling.  
In the fitting process, when a model parameter deviates 
from it's optimal value the other parameters try to adjust themselves 
so as to keep the $\chi ^2$ value as small as possible.  This situation 
not only exists when fitting an $r^{1/n}$ bulge model, but also when
fitting an $r^{1/4}$ or an exponential bulge model.  One way to gauge the 
extent of such `coupling' could be to vary one parameter, while allowing 
the other parameters to compensate, until the $\chi ^2$ value
increases by some fixed amount.  This was the approach taken
in Graham et al.\ (1996), where it was shown that coupling of the $r^{1/n}$ 
model parameters in fits to brightest cluster galaxies (ellipticals) is not
responsible for the trend between profile shape $n$ and galaxy size $r_e$.  
However, as was noted before, for spirals with relatively small bulges, the 
global $\chi ^2$ statistic can be rather insensitive to changes in the bulge
model and so it is not an appropriate quantity to use here.  One could try 
using the 
$\chi ^2$ value within 2$r_{b=d}$, but to do this one would have to fit
the bulge and disk models to this radial range alone.  In so doing one would no 
longer be measuring the best global solution which is what one wants. 
It's a little like fitting a model nose and body to the silhouette 
of an elephant.  The goodness of the fit is dominated by the model for the body
irrespective of whether you fit an ear or a nose to what should be the nose.
That is, one may fit significantly different shapes for the nose, while the 
global $\chi ^2$ value changes little.  In an attempt to better 
gauge the goodness of the fit to the nose, one could zoom in, changing all the
model parameters, and try to 
increase the quality of the fit in just the area around the nose -- but this 
would come at the expense of a good fit for the body.  One can, and should, 
fit both the nose (bulge) and body (disk) simultaneously to find the best 
fit (Schombert \& Bothun 1987).  Although, when one does this, the
goodness-of-fit in the area around the nose certainly does not have to 
be the best fit for that specific area, and different model parameters may
well give a better fit in that region, albeit, at the expense of the global fit.
Despite all this, when the shape parameter $n$ was changed from the optimal 
solution to the traditional values of 1 and 4, Figure~\ref{ChiRat} reveals 
that any possible coupling between the model parameters was not sufficiently 
large as to prevent significant increases to the $\chi ^2$ values over those
obtained with an $r^{1/n}$ bulge and an exponential disk model. 

The global fits are shown in Figure~\ref{figApp}, where the models are 
represented by the dashed lines, and the seeing-convolved models are 
represented by the 
solid lines, which, in most instances, predominantly over-lap each other. 
While an exponential model fits the disks extremely well in most cases, there 
are a few exceptions besides the galaxy UGC 6028, which, as mentioned before, 
has been excluded.  This can be seen, for example, in the profile residuals 
of UGC 05842 and UGC 12614.  These two galaxies possess disks which appear 
to have a shape parameter $n$$<$1.  In fact, about a dozen galaxies possessed 
disks which deviated from a straight exponential light profile.  These 
departures may be real for the 
galaxies: UGC 00490, UGC 06453, UGC 07169, UGC 07901, and UGC 00242.
While for other galaxies it may simply be due to errors in the estimate of
sky background level which cause the outer few data points to depart from 
an exponential profile.  

\begin{figure}
\centerline{\psfig{figure=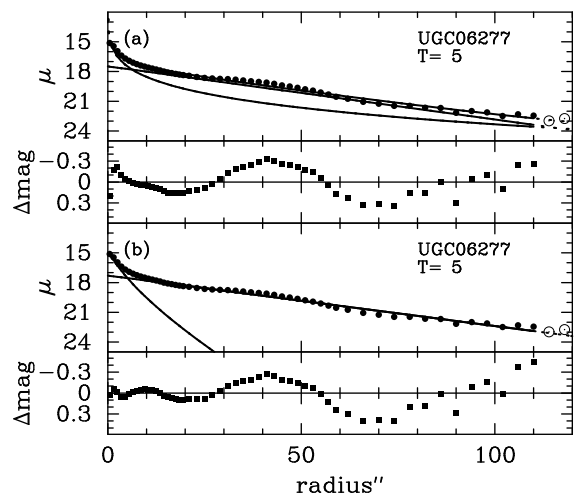,width=7.5cm,angle=-90}}
\caption{This plot illustrates, with the $K$-band data from UGC 06277,
that secondary minima
can exist in the ocean of $\chi ^2$ values resulting from different light profile
model parameters.  In the upper panel (a) the $\chi ^2$ value is 1.70, and in the
lower panel (b) $\chi ^2$=1.63.  In panel (a), the use of poor initial estimates
for the model parameters resulted in the minimization code getting trapped in a
local minima and leading to a poor match to the shape of the bulge profile.
The dotted curves are the models, while the solid curves are the seeing convolved
models which are fitted to the data.}
\label{kick}
\end{figure}

\begin{figure}
\centerline{\psfig{figure=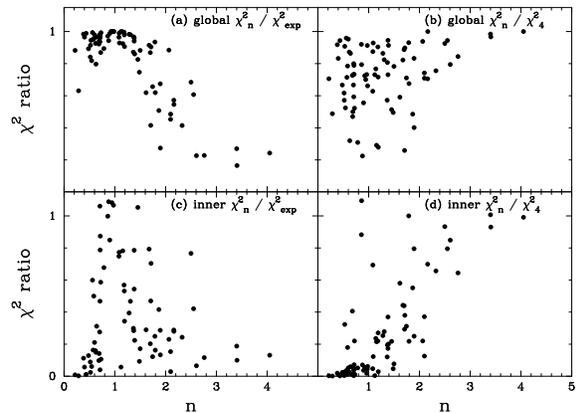,width=7.5cm,angle=-90}}
\caption{The $\chi ^2$ values from the $r^{1/n}$, $r^{1/4}$, and exponential
bulge models, fitted simultaneously with an exponential disk model to the
$K$-band light profiles, have been given the subscripts $n$, 4, and exp.
The upper panels (a) and (b) show the `global' $\chi ^2$ ratios fitted to the
entire light profile, while the lower panels (c) and (d) show the $\chi ^2$
ratios from the profile within the two times the radius where the bulge and
disk light contribute equally.  The same best-fitting model parameters, as
determined from the global fit, have been used in both the upper and lower
panels.}
\label{ChiRat}
\end{figure}

To avoid such data points which may be wrong due 
to an incorrect sky-background estimate, suspect points have been marked in 
Figure~\ref{figApp} with a circle and excluded from the fitting procedure.  
In some cases, not performing this truncation actually resulted in the bulge 
model kicking up and contributing more light to the outer portions of the 
galaxy than the disk.  This situation, illustrated in Figure~\ref{kick} 
arises because of the failure of the exponential disk model 
to allow for curvature in the outer profile.  The bulge, able to account 
for such curvature, comes into play by adding light to the disk in the 
outer parts of the galaxy and reducing the $\chi ^{2}$ value of the 
fitted models.  Such fits could be telling 
us that the bulges of some spiral galaxies are a lot more extended and 
prevalent than previously thought; although, this seems unlikely as the
spiral structure usually dominates the light from the outer parts of the images. 
This example also serves to illustrate the point that secondary minima 
exist within the ocean of $\chi ^{2}$ values arising from all the different 
model parameter combinations.  Depending on the initial 
estimates which one provides for the model parameters, one may end up in one 
of a number of unfavorable minimum.  However, while this occurred for 10-20\% of
the sample after the first run through, a visual inspection of the model fits 
and the profile data easily revealed when the program had converged on an 
undesirable minimum (as in Figure~\ref{kick}).  For those galaxies which were
sensitive to poorly matched initial estimates for the model parameters, the
code was run again with initial estimates that more closely resembled the 
galaxy profile.   The results are presented in Figure~\ref{figApp}. 
% where one can judge for themselves whether or not the code converged on 
% an inappropriate minimum.  
This situation highlights the fact that one cannot 
always be passive and blindly run their minimization codes faithfully beleiving 
that they will always find the optimal physically realistic solution.  However, 
one can increase their chances by using smarter codes which can climb their 
way out of local minimum to locate the global minimum.

%\placefigure{kick}

The $\chi ^2$ statistic has been used as an estimator for the quality of 
the fit.  However, a variant of what might be considered the usual practice 
has additionally been employed.  For a number of the galaxies, compared to 
their disk, their bulge 
appears relatively small and the global $\chi ^2$ value is dominated by the 
fitted exponential disk, largely irrespective of the bulge model.  Clearly, in 
such circumstances, the global $\chi ^2$ statistic is largely insensitive to the
bulge model.  Therefore, to estimate the goodness-of-fit of the bulge and disk
model, the $\chi ^2$ values within two times the radius where the bulge
and disk light contribute equally ($r_{b=d}$, see Section~\ref{EqLuz}) 
were measured.  For three galaxies whose bulge light was less than the 
disk light even at the center of the galaxy, the inner 3$\arcsec$ radius 
was used.  It should be noted that the bulge and disk model have 
still been fitted to minimize the global $\chi ^2$ value, and not the 
$\chi ^2$ value within 2$r_{b=d}$.  As a consequence, in some instances 
use of the $r^{1/n}$ bulge and exponential disk model may result in a 
larger $\chi ^2$ value than the classic bulge models over this radius.
However, in general this is not the case as is illustrated in 
Figure~\ref{ChiRat}, which plots the ratio of the $\chi ^2$ values 
from the different models as a function of $n$.  One can see that both
globally, and over the inner 2$r_{b=d}$, the $r^{1/n}$ bulge model 
results in a better fit than the fixed $n$$=$1 or 4 models.  

To inspect whether the fits from the use of an $r{1/n}$ bulge provide, on 
average, a statistically significant improvement over the fits from the 
use of the classical bulge models, Student's t-test was applied to the 
distributions of $\chi ^2$ values.  The fits with the $r^{1/n}$ bulge
models gave significantly (99.5\%) smaller $\chi ^2$ values -- within 
2$r_{b=d}$ -- than the fits with an exponential bulge model, and smaller 
still values (at the 99.98\% level) than those obtained when using the 
classical $r^{1/4}$ bulge and exponential disk model.

%\placefigure{ChiRat}

\begin{figure}
\centerline{\psfig{figure=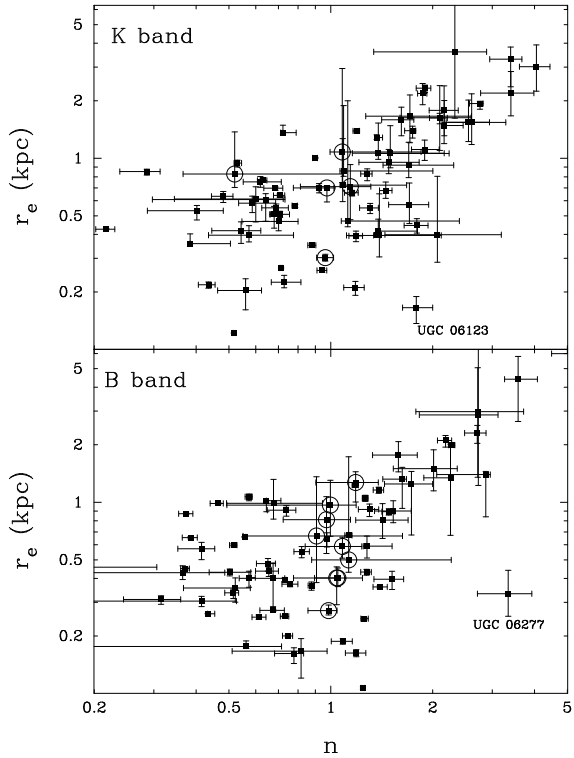,width=7.5cm,angle=00}}
\caption{The bulge effective half-light radius $r_e$ is plotted against the
S\'{e}rsic $r^{1/n}$ bulge shape parameter $n$.  The errors are from the
uncertainty in the sky-background level.  Those data points with circles
around them are consistent with a value of $n$=1.}
\label{err1-fig}
\end{figure}

\begin{figure}
\centerline{\psfig{figure=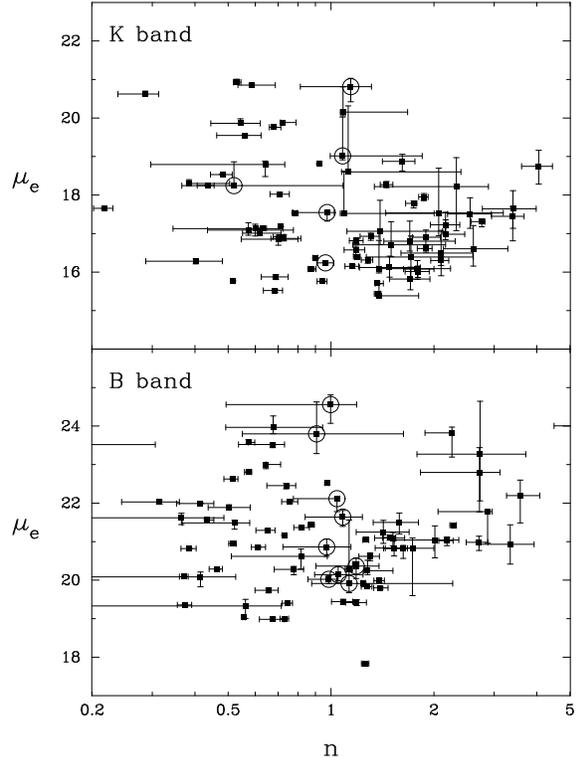,width=7.5cm,angle=00}}
\caption{The surface brightness, $\mu_e$, of the bulge profile at $r_e$, is
plotted against the S\'ersic $r^{1/n}$ bulge shape parameter $n$.  The
errors are from the uncertainty in the sky-background level.  Those data
points with circles around them are consistent with a value of $n$=1.}
\label{err2-fig}
\end{figure}

Figure~\ref{err1-fig} reveals that not only is there a positive
correlation between $n$ and the bulge effective radius (a result also 
seen in Khosroshahi et al.\ 2000;
their figure 3), but that most of the bulge luminosity profiles are not consistent 
with an exponential light distribution.  In the $K$-band, only 5 galaxies have a 
bulge shape parameter $n$ that is consistent with the exponential value of 1.  
Figure~\ref{err2-fig} shows the distribution of $n$ with the effective bulge 
surface brightness.

%\placefigure{err1-fig}

%\placefigure{err2-fig}

\begin{figure}
\centerline{\psfig{figure=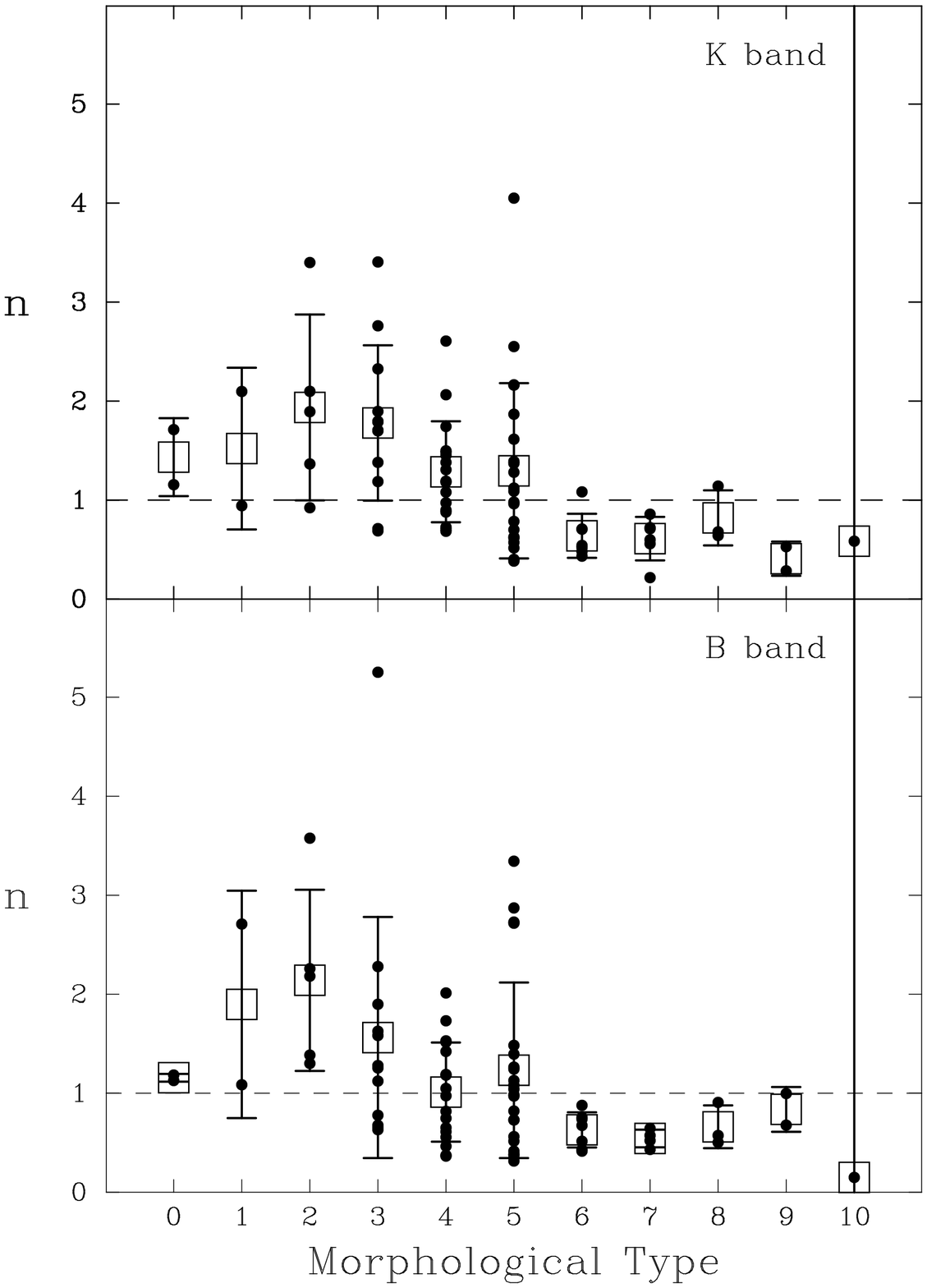,width=7.5cm,angle=00}}
\caption{For each galaxy, the best-fitting S\'{e}rsic $r^{1/n}$ bulge shape
parameter $n$ is plotted against the galaxy's morphological type index.}
\label{nT-fig}
\end{figure}

In Figure~\ref{nT-fig} a trend can be seen between galaxy type, 
or numerical stage index $T$ (de Vaucouleurs et al.\ 1991), and the 
light-profile shape of a spiral galaxy's bulge. 
The early-type spiral galaxies have bulges which are better described 
with shape parameters $n$$>$1, while late-type spiral galaxy bulges are
better described with shape parameters $n$$<$1.  This result expands --
into the domain of late-type galaxies -- upon the trend of decreasing $n$
with increasing $T$ which was observed by Andredakis et al.\ (1995; their 
figure 5a) for disk galaxies with morphological type $T$$\leq$5.  
%
% Similarly, all of the 26 early-type disks modelled by Khosroshahi 
% et al.\ (2000) had bulge shape parameters larger than, and inconsistent 
% with, a value of 1.   
% 
% In passing, it is noted that they used galaxies from a high-inclination 
% sample, where use of the spiral arm structure to determine morphological type 
% is more difficult. 
%
The shape parameter $n$ is therefore more than just an additional parameter 
that improves the quality of the fitting routines, but traces a physical 
characteristic of the bulges of disk galaxies. The inner light profiles of
late-type spiral galaxy bulges are flatter than those in early-type spiral
galaxy bulges and decline more quickly with radius in the outer parts. 

\begin{figure}
\centerline{\psfig{figure=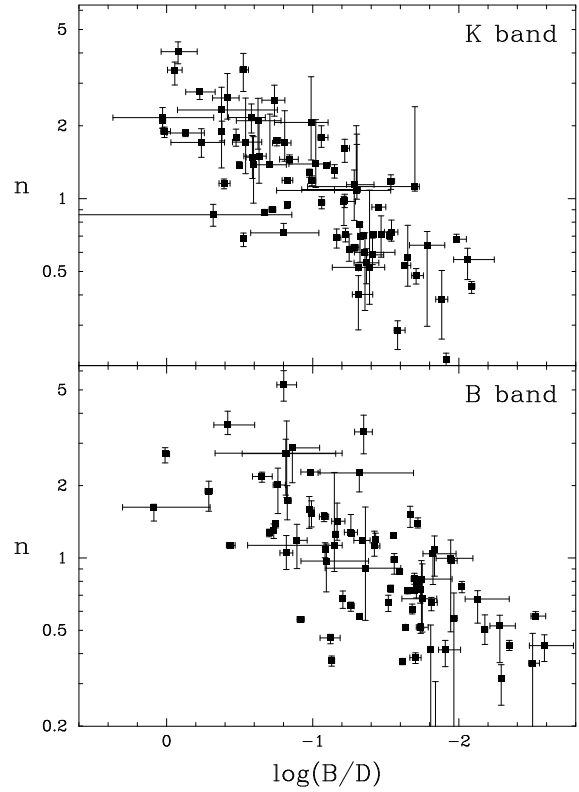,width=7.5cm,angle=00}}
\caption{For each galaxy, the best-fitting S\'{e}rsic $r^{1/n}$ bulge shape
parameter $n$ is plotted against the logarithm of the bulge-to-disk luminosity
ratio.}
\label{n-BD-fig}
\end{figure}

A plot of $n$ against the $B/D$ luminosity ratio is shown 
in Figure~\ref{n-BD-fig}.  In the $K$-band, the Pearson correlation coefficient 
between $n$ and $\log(B/D)$ is 0.75, and the Spearman rank-order correlation 
coefficient is 0.80, with the two-sided significance level of its deviation from
zero less than $10^{-18}$ (that is, this correlation is highly significant).  
In the $B$-band these values are 0.60 and 0.70 respectively, with similarly
high significance.  The current extension to later type spiral galaxies supports
and strengthens the correlation between $n$ and the $B/D$ luminosity ratio seen 
in Andredakis et al.\ (1995), where they reported a linear correlation
coefficient of 0.54 at a significance level of 99.7\% for their $K$-band 
sample of early-type disk galaxies.

%\placefigure{nT-fig}

%\placefigure{n-BD-fig}

\begin{figure}
\centerline{\psfig{figure=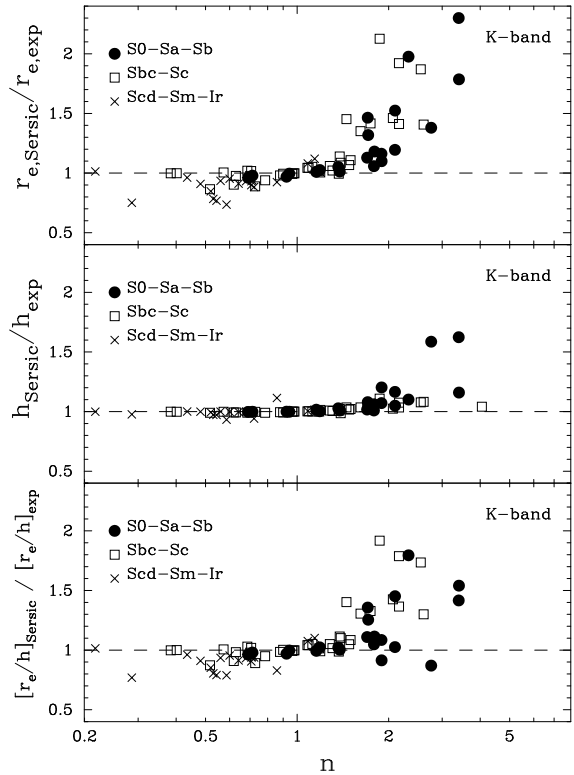,width=7.5cm,angle=00}}
\caption{A comparison plot of the bulge effective half-light radii $r_e$, the
disk scale-length $h$, and the ratio of these quantities $r_e/h$, between the
estimates using the best-fitting S\'{e}rsic $r^{1/n}$ bulge versus an exponential
bulge.  An exponential disk was simultaneously fitted in both instances.}
\label{old-new1}
\end{figure}

\begin{figure}
\centerline{\psfig{figure=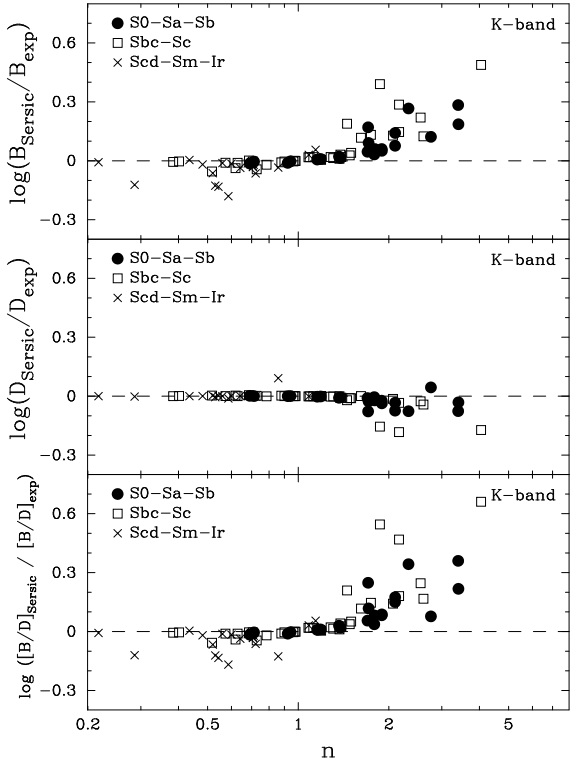,width=7.5cm,angle=00}}
\caption{A comparison plot of the bulge luminosity $B$, the
disk luminosity $D$, and the ratio of these quantities $B/D$, between the
estimates using the best-fitting S\'{e}rsic $r^{1/n}$ bulge versus an
exponential bulge.
An exponential disk was simultaneously fitted in both instances.}
\label{old-new2}
\end{figure}

The data were re-modelled using a seeing-convolved exponential bulge 
model and a seeing-convolved exponential disk model.  
In Figure~\ref{old-new1} the effective bulge radius derived from fitting 
a S\'{e}rsic bulge is plotted against the effective bulge radius derived 
from fitting an exponential bulge.  One can 
clearly see that fitting an exponential law to the bulge of a 
spiral which has a S\'{e}rsic index greater than 1 will lead to an 
under-estimation of the true effective bulge radius.  On the other hand, if 
the bulge light profile shape is better described with a S\'{e}rsic index less 
than 1, fitting an exponential bulge will result in over-estimation of the 
effective bulge radius.  Even when just considering galaxies with bulge 
values of $n$ ranging from 0.5 to 2.0, the range of $r_e$ computed using 
an exponential bulge is a factor of 2 less than the real range. 
Changes to the $r_e/h$ ratio are dominated by changes in $r_e$ rather than 
changes in the disk scale-length $h$. 
Differences to the bulge-to-disk luminosity ratio are shown in 
Figure~\ref{old-new2}, where it can be seen that use of an exponential
bulge model systematically under-estimates the bulge luminosity when the 
bulge profile shape has $n$$>$1, and subsequently results in an under-estimation
of the bulge-to-disk luminosity ratio.  When $n$ is smaller than 1, the 
exponential bulge model results in an over-estimation of both the bulge 
luminosity and the bulge-to-disk luminosity ratio.

%\placefigure{old-new1}

%\placefigure{old-new2}

\begin{figure}
\centerline{\psfig{figure=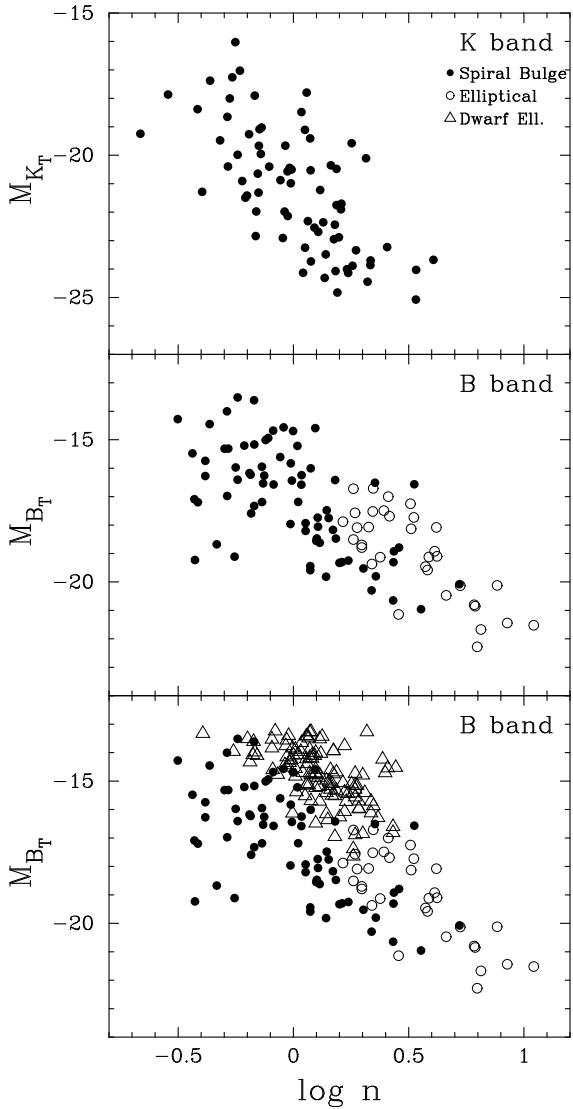,width=7.5cm,angle=00}}
\caption{The total absolute magnitude of the spiral galaxy bulges is
plotted against the logarithm of the best-fitting S\'{e}rsic shape parameter.
For comparison, a sample of Virgo dwarf elliptical (dE) galaxies from
Jerjen Binggeli, \& Freeman (2000) is shown, as are a sample of `ordinary'
Virgo and Fornax Elliptical (E) galaxies from Caon et al.\ (1993) and D'Onofrio
et al.\ (1994).  A Hubble constant of 75 km s$^{-1}$ Mpc$^{-1}$, and
a Virgo distance-modulus of 31.05 have been used.}
\label{fig-todo}
\end{figure}

Figure~\ref{fig-todo} shows that the total bulge light (in both the $B$- and $K$-band) 
-- and therefore possibly bulge mass -- also correlates strongly with the shape of 
the bulge light profile (Pearson's $r$$=$0.68, Spearman's $r_s$$=$0.69). 
Such behaviour is also seen in the bulge data of Andredakis et al.\ (1995; their
figure 6b). 
Systematic trends in the shape of the light profile with luminosity have
been seen before amongst the dwarf elliptical population (Caldwell \& Bothun 
1987; Binggeli \& Cameron 1991) and parametrized with the S\'{e}rsic model 
(Davies et al.\ 1988; Young \& Currie 1994, 1995; Jerjen \& Binggeli 1997; 
Jerjen, Binggeli, \& Freeman 2000).  Similarly, this same behaviour of increasing
S\'{e}rsic shape parameter $n$ with absolute luminosity has been seen amongst the 
elliptical galaxy population (Michard 1985; Schombert 1986) and subsequently 
parametrized (Caon, Capaccioli, \& D'Onofrio 1993; D'Onofrio, Capaccioli, 
\& Caon 1994; Hjorth \& Madsen 1995; Graham \& Colless 1997, Gerbal et al.\ 1997;
Lima Neto 1999).  Indeed, this pattern has also been observed amongst the 
bulges of S0 galaxies (Capaccioli 1987, 1989).  
It is therefore of interest to compare all of these objects on the one diagram to 
look for similarities and differences.  The lower panel of Figure~\ref{fig-todo} 
is an extension of the plot shown in Jerjen \& Binggeli (1997, their figure 2). 
The dwarf Elliptical (dE) photometry is from the Virgo dEs given in Binggeli \& 
Cameron (1991, 1993) -- most recently presented in Jerjen, Binggeli, \& Freeman 
(2000, their figure 6).  A Virgo distance-modulus of 31.05 mag has been used 
(Jerjen, Freeman, \& Binggeli 2000).  The `ordinary' elliptical galaxy data set 
are from the Virgo and Fornax Ellipticals presented in Caon et al.\ (1993) and 
D'Onofrio et al.\ (1994), excluding 3 galaxies, from a total of 38, which were 
labelled as 
having a `poor' profile fit, and excluding the one E,pec galaxy.  A Virgo-Fornax 
distance-modulus of 0.25 mag was used (Graham 1998, and references within). 

Viewed on its own, the middle panel of Figure~\ref{fig-todo} appears to show 
that the bulges of spiral galaxies form the faint extension to the elliptical 
galaxies.  However, the location of the dwarf ellipticals in the lower panel
reveals that this picture is not so clear.  Given that, at least structurally, 
the dwarf elliptical galaxies are the smaller, fainter counterparts to 
the brighter ellipticals (Caon et al.\ 1993; Graham et al.\ 1996; Jerjen 
\& Binggeli 1997), the bulges of spirals show a distinctly different 
behaviour to the ellipticals.  For a given 
luminosity profile shape, i.e.\ $n$, spiral bulges are brighter than ellipticals
with the same light profile shape, or alternatively, for a given luminosity,
spiral bulges have a shallower core distribution of stars and a steeper fall-off
at larger radii (i.e.\ smaller $n$).  Perhaps the rotating disk is responsible
for truncating the bulge and creating the smaller $n$ parameters.  However, 
what is the cause and what is an affect is not clear.  What is clear is that 
the abandonment of the limited classical fitting functions has given rise to 
a potentially powerful diagnostic tool.  

% The location of the bulges of S0 galaxies in the $n$-$log(B/D)$
% diagram may indicate if they more akin, at least structurally, to the bulges 
% of spiral galaxies or if they are more like elliptical galaxies.  
% The bulges of the S0 galaxies from the $K$-band images of Andredakis et al.\ (1995) 
% have shape parameters 2$<$$n$$<$6.  These estimates have been obtained while 
% simultaneously modelling for the impregnated disk, and do not represent the value
% of $n$ fitted to the entire light profile.  

%\placefigure{fig-todo}

\section{Analysis of the bulge-to-disk ratios with morphological type}

\subsection{Bulge-to-disk size ratio}

Graham \& Prieto (1999a) showed that the universal application of an 
exponential light-profile model to the bulges of spiral galaxies 
produced a mean $r_e/h$ value that was actually smaller for the early-type
galaxies than for the late-type galaxies -- at the 98\% 
significance level using the $K$-band data of de Jong (1996a). 
This result was clearly at odds with the popular conception of what 
early- and late-type spiral galaxies look like and brought us to examine the
use of the exponential bulge model for drawing conclusions about the
nature and structural properties of spiral galaxies. 

Fitting an $r^{1/n}$ bulge -- which allows for the range of structural
profile shapes that spiral bulges possess -- the mean value of $r_e/h$ 
for the early- and late-type spirals from de Jong's sample of `face-on' 
galaxies is re-derived.  Table~\ref{stats-r-h} gives the probability 
that the early- and late-type spirals have the same mean $r_e/h$ value 
depends only mildly on the passband used; the bluer passbands yielding the
greatest distinction between the mean $r_e/h$ value for the early- and late-types. 
Although, in general, the mean $r_e/h$ value is larger for the early-type 
spirals than the late-type spirals, the difference is not particularly significant. 
The increased probability in Student's t-test when certain galaxies are excluded 
is, to a degree, because of the reduced number of galaxies and consequent lack 
of strength in the statistical test.  Although, the presence of a bar does 
appear to bias the mean $r_e/h$ value for the early-type spirals.  de Jong 
(1996a) found that he obtained a better fit for 23 of the 86 spiral galaxies 
when he included an additional bar component in the 2D analysis.  After the 
removal of those 23 barred galaxies, the mean $r_e/h$ values are in agreement 
with each other.

\placetable{stats-r-h}

The robustness of this result is investigated against further modifications 
to the actual sample used.  Inclusion of the few S0 and Irregular type galaxies 
did not change the probabilities by more than a few percent (and by no more 
than 1\% for those probabilities less than 10\%). 
Exclusion of the Scd galaxies from the late-type sample changed the 
mean $r_e/h$ value by only 0.002-0.006, but the smaller galaxy numbers in the 
sample meant that the strength of Student's t-test was reduced and the 
probabilities increased because of this.

\subsection{Bulge-to-disk luminosity ratio}

The bulge-to-disk luminosity ratio is commonly thought to be a fundamental 
characteristic to the Hubble sequence, and by this it is meant the
revised Hubble sequence of Sandage (1961), such that the prominence of the
bulge decreases with the later spiral galaxy types.  Sandage had however 
changed the 
defining criteria to that of the spiral arms, and so to investigate the 
above belief, the present galaxy sample was again separated into early- 
and late-types and Student's t-test employed to measure the probability 
that the two $\log(B/D)$ luminosity ratio distributions have different mean values. 
Table~\ref{stats-B-D} shows that the probability that the mean $\log(B/D)$
luminosity ratios could be as
different as they are just by chance is less than 0.3\% (3$\sigma$). 
The F-test reveals that the two $\log(B/D)$ ratio distributions have similar
variances, and so only the results from Student's t-test with similar 
variances is shown. 

\begin{figure}
\centerline{\psfig{figure=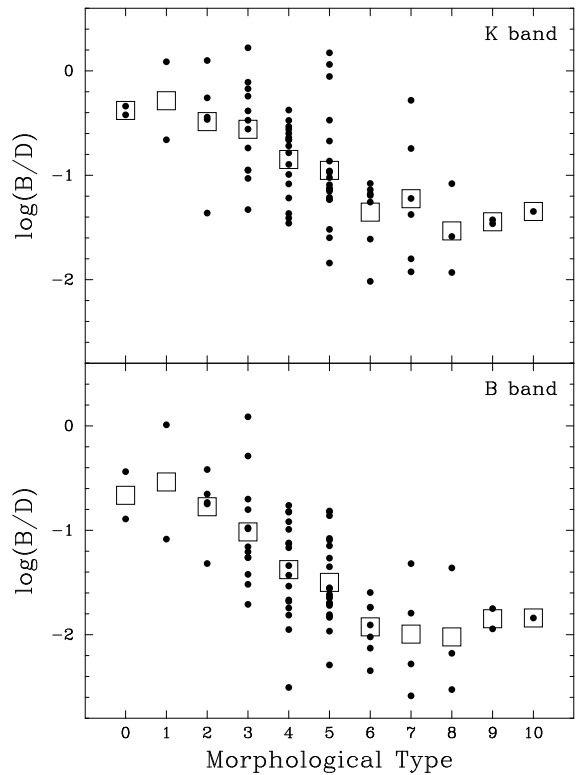,width=7.5cm,angle=00}}
\caption{The logarithm of the S\'{e}rsic bulge to exponential disk luminosity,
as given in equation~\ref{MuEq}, is plotted against morphological type.
The disk luminosity has been corrected for inclination effects as described
in the text.  The squares mark the mean value for each galaxy type.}
\label{BD-fig}
\end{figure}

While the $r_e/h$ ratio, and similarly $(r_e/h)^2$, are not responsible for 
the significantly different mean $\log(B/D)$ ratios from the early- and late-type 
spiral galaxy sets, the other two components to equation~\ref{MuEq} are.  
That is, both the mean difference in the surface brightness term $\log(I_e/I_0)$ 
and the structural differences given by $n$ are different for the two samples,
and responsible for the $B/D$ luminosity ratio decreasing with increasing $T$-type.
The logarithm of the $B/D$ luminosity ratio has been plotted as a function of 
galaxy type in 
Figure~\ref{BD-fig}.  Within each $T$-type, while the spread of values can be 
large, the mean value steadily decreases from type $T$$=$1 (Sa) to $T$$=$8 (Sdm). 
Although, from the $\log(B/D)$ ratio alone it is not possible to state with 
confidence the morphological type; a conclusion also reached by Simien \& 
de Vaucouleurs (1986) and evident in Figure~\ref{amazing}.  Perhaps this isn't 
surprising given that the first criteria of galaxy classification refers to 
the nature (pitch angle and resolution) of the spiral arms -- which must have 
a somewhat similarly loose correlation with the $\log(B/D)$ ratio. 
The greater degree of scatter in the $K$-band is likely to be tied in with how the 
pitch-angle and spiral-arm nature change once the obscuring dust `mask' is
penetrated at near-infrared wavelengths (Block \& Puerari 1999), and the 
optical Hubble classifications are no longer entirely appropriate.

\placetable{stats-B-D}

%\placefigure{BD-fig}

\section{Other quantitative galaxy classification parameters of bulge strength}

\subsection{The concentration index}

% Using $r^{1/4}$ bulge models, Schombert \& Bothun (1987) urged that the $B/D$ 
% luminosity ratio be abandoned in favour of light-concentration indices. 

Morgan \& Mayall (1957) found the spectroscopic characteristics 
(stellar populations)
from the inner parts of a galaxy correlated well with the central concentration 
of luminosity.  Noting the lack of agreement between the spiral arm structure 
and the nuclear region for many galaxies, Morgan (1958, 1959, 1962) excluded 
the former criteria from his qualitative classification scheme which was based 
on the central concentration of light (known as the Yerkes system). 
de Vaucouleurs (1977) made this method of classification quantitative with the 
introduction of the concentration index $C_{31}$, defined as the ratio between 
the radii that contain 75\% and 25\% of the total luminosity.  Okamura, Kodaira, 
\& Watanabe (1984) showed that the different concentration indices one could
construct are essentially the same, and so variations came into use.  For example, 
Kent (1985) used the 20\% and 80\% radii, while Gavazzi, Garilli, \& Boselli 
(1990) replaced the total magnitude with the $V$ magnitude enclosed by the 
25 mag arcsec$^{-2}$ isophote.  

Doi et al (1992) and Doi, Fukugita, \& Okamura (1993) suggested the use of
the concentration index for the automatic classification of nearby, and small 
and faint, galaxies rather than as a means to characterise morphological type.  
Since then, a small industry has been established (Abraham et al.\ 1994; 
Fukugita et al.\ 1995) and the concentration index has become a popular 
diagnostic for high redshift galaxies, where the nature of the spiral arms is
less clear.  It's use has continued for studies 
of nearby galaxies, most recently in the near-infrared, where Moriondo et 
al.\ (1999) used $C_{31}$ at $H_{21.5}$, and Gavazzi et al.\ (2000) used 
$C_{31}$ at $H_{T}$.  In this paper the $K$-band value for 
$C_{31}$ is derived within the $K_{22.0}$ mag arcsec$^{-2}$ isophote, and the 
$B$-band value for $C_{31}$ is derived using the $B_{25.0}$ mag arcsec$^{-2}$ isophote.
The value for $C_{31}$ was also derived using the total galaxy light, but this
had little difference on the overall trends and so is not presented here. 

\begin{figure}
\centerline{\psfig{figure=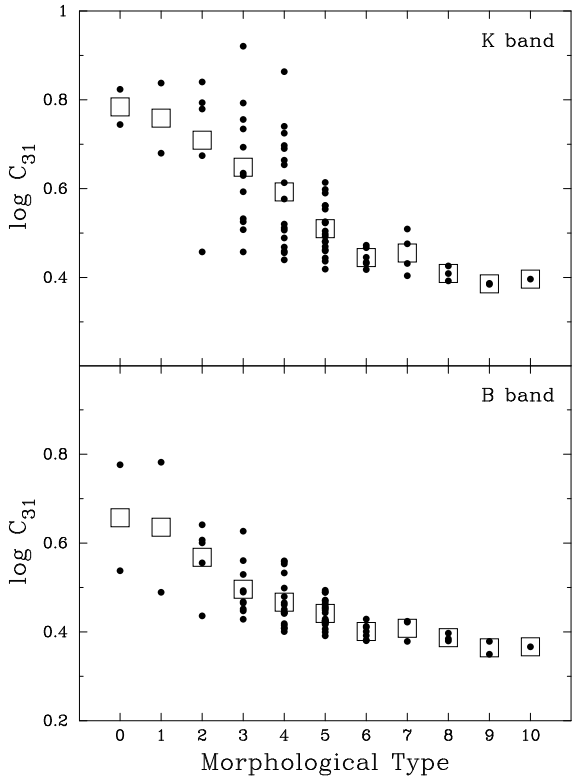,width=7.5cm,angle=00}}
\caption{The logarithm of the concentration index $C_{31}$, as defined in the text,
is plotted against morphological type.}
\label{C-T-fig}
\end{figure}

Figure~\ref{C-T-fig} reveals, not surprisingly, a somewhat similar behaviour 
to that seen in Figure~\ref{BD-fig}, but with perhaps slightly less scatter.
It would appear that one cannot use the concentration 
index, or the $B/D$ ratio, to determine a galaxy's morphological type -- an
issue taken to heart by Abraham (1999), where it is suggested, at least for
studies of distant galaxies, that morphological type be replaced by more 
quantitative measures   A similar behaviour to that seen in Figure~\ref{C-T-fig} 
is evident in the $H$-band data set of Moriondo et al.\ (1999), where the later 
type spiral galaxies, from a sample of nearby galaxies, all tend to have low 
concentration indices around 0.4 -- as expected for bulgeless disks -- 
and several of the early-type spiral galaxies also have low concentration 
indice values. 

\begin{figure}
\centerline{\psfig{figure=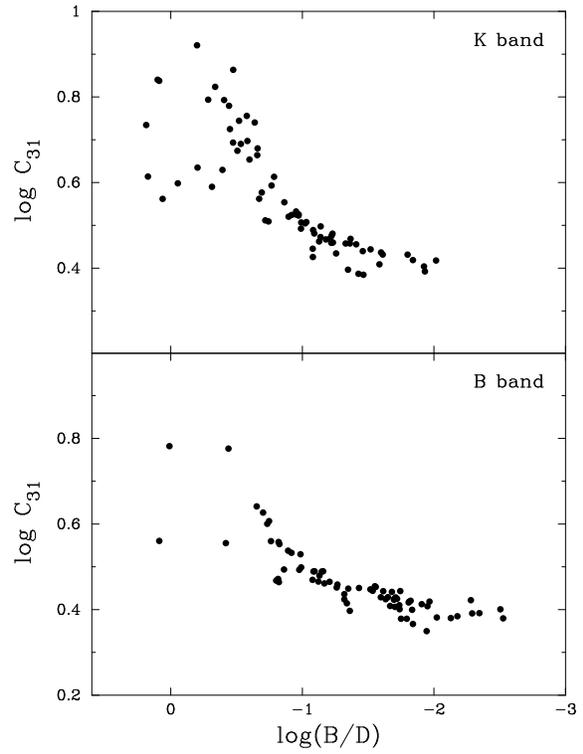,width=7.5cm,angle=00}}
\caption{The logarithm of the concentration index $C_{31}$, as defined in the text,
is plotted against the logarithm of the bulge-to-disk luminosity ratio.}
\label{C-BD-fig}
\end{figure}

The logarithm of the concentration index has been plotted against the logarithm 
of the $B/D$ luminosity ratio in Figure~\ref{C-BD-fig}.  The scatter in the
relationship along the lower right of the curve is because $C_{31}$ has been 
computed using the total galaxy magnitude within some limiting isophote, 
rather than extrapolating to infinity as was done with the $B/D$ ratio.   
The increased scatter for $B/D$ ratios greater than about $\sim$0.3 arises
because sometimes the bulge light is no longer concentrated within the 
radius containing 25\% of the total galaxy light.  Consequently, the concentration
index is not a good tracer of the logarithm of the $B/D$ luminosity ratio
when the bulge is relatively large and extended in comparison to the disk. 
Use of the 40\% and 80\% radii improved things a little at the high $B/D$ ratio end 
of this relation, but at the expense of sensitivity at the low $B/D$ ratio end -- 
that is, most galaxies with $B/D$ ratios less than 0.1 ended up having roughly 
the same concentration index.

%\placefigure{C-T-fig}

%\placefigure{C-BD-fig}

\subsection{Two new parameters to measure the apparent prominence of the bulge\label{EqLuz}}

Certainly not rejecting the concentration index at this point, this section
does however explore other quantitative estimators of bulge strength.  
While the $r_e/h$ ratio may be useful for galaxy modellers, it apparently does 
not correlate 
strongly with the assigned morphological type of a galaxy, nor does it necessarily 
reflect the `apparent' prominence of the bulge.  Due to the over-lapping of 
the bulge and disk, what the eye sees (which is what has been used to classify 
galaxy type) can be mis-leading and has given rise to this apparent contradiction.  
Therefore, two new quantitative estimators of bulge strength which 
more accurately reflect what the eye discerns when it looks an an image of a 
spiral galaxy are introduced.  This should allow us to test if the galaxy 
sample is perhaps unusual and not representative of the larger galaxy population. 

Firstly, the difference between the observed central surface brightness of 
the galaxy and the surface brightness where the bulge and the disk contribute 
equally was used as a quantitative measure for the prominence of the bulge.  
Although the observed central surface brightness will be affected by seeing, 
because morphological classification is also done with seeing affected images this 
is the value that has been chosen.  There was another motive for using this value
rather than the seeing-corrected value, even though this latter quantity is the
more appropriate one to use.   The reason was to make the result 
independent of any fitted light profile model. 

The situation is a little more complicated with the $K$-band data.  Due to the 
transparent nature of the disk, the apparent radius where the bulge and disk 
light contribute equally will actually be dependent on the inclination of the
disk, and so the disk surface brightness should first be corrected to some 
standard value -- corrections to the face-on value were applied.  Due to this 
effect, in the $K$-band, one would expect the observed prominence of the 
bulge to be less for a sample of highly inclined galaxies than
a sample of low-inclination galaxies.  Additionally, this will make near-infrared 
classifications which are based on the prominence of the bulge prone to error.  

\begin{figure}
\centerline{\psfig{figure=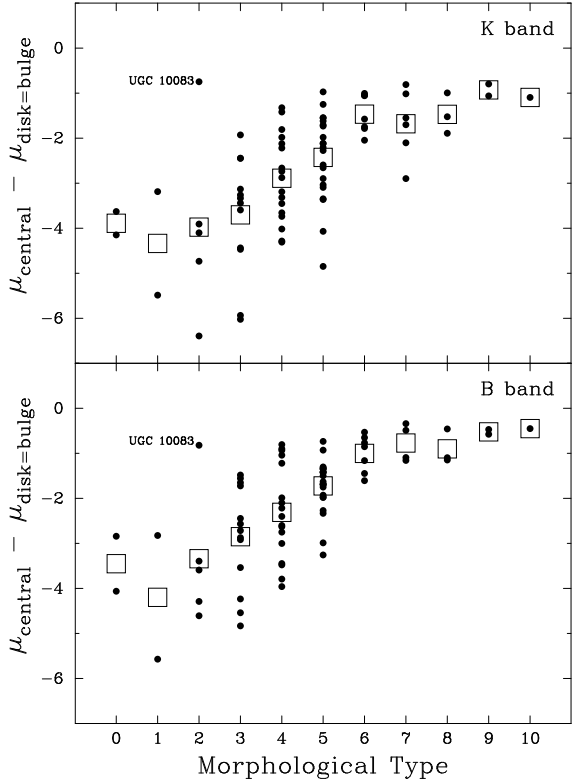,width=7.5cm,angle=00}}
\caption{The difference between the central galaxy surface brightness
and the surface brightness where the bulge and disk light contribute equally
is plotted against galaxy morphological type.  The effects of seeing are
deliberately not taken into account.}
\label{SB2a-fig}
\end{figure}

The average difference between the central galaxy surface brightness and the 
surface brightness where the bulge and disk light contributes equally was 
computed for both the early-type ($T$$\leq$3) and late-type ($T$$\geq$6) 
spiral galaxies, and the individual data for all galaxies is shown in 
Figure~\ref{SB2a-fig}.  Application of Student's t-test (Table~\ref{Stats-Mu}) 
revealed the probability that the early- and late-type samples could have the 
same mean difference can be ruled out 
at greater than the 2$\sigma$ significance level.  This is not a result of the 
S\'{e}rsic models with large $n$ having steeply rising light profiles at their 
centers; this result is completely independent of the profile models.  If we
had of used the seeing-corrected, model dependent, central surface brightness 
values, then this result would be even stronger.  That the early-type spiral 
galaxies have, on average, a greater difference between the central surface 
brightness and the surface brightness where the bulge and disk are equally 
bright, is in accord with 
expectations, and so it seems probable that the galaxy sample is not peculiar
or biased. This gives some reassurance that the $r_e/h$ results presented earlier 
are likely to be accurate for the population as a whole.

%\placefigure{SB2a-fig}

From the plot of $\mu _{\rm central}-\mu _{\rm bulge=disk}$ one point stands out in 
Figure~\ref{SB2a-fig}.   The galaxy UGC 10083, catalogued as (R)SB(r)ab ($T$$=$2)
in NED, does not have an obvious bulge and has a small value for 
$\mu _{\rm central}-\mu _{\rm bulge=disk}$. 
Inspection of it's image and profile suggest that this galaxy may be more 
like an Sbc or Sc galaxy (2 or 3 Hubble T types from it's catalogued value; 
which is perhaps not unreasonable, Lahav et al.\ 1995).
To use Figure~\ref{SB2a-fig} again as a diagnostic tool, it is noted that 
NED gave UGC 09024 the morphological type `S?'.  It has a value for 
$\mu _{\rm central}-\mu _{\rm bulge=disk}$ of around 5.3 mag arcsec$^{-2}$ 
($K$-band) and 4.0 mag arcsec$^{-2}$ ($B$-band), suggesting it's type index 
is probably around $T$$=$3$\pm$2. 

\begin{figure}
\centerline{\psfig{figure=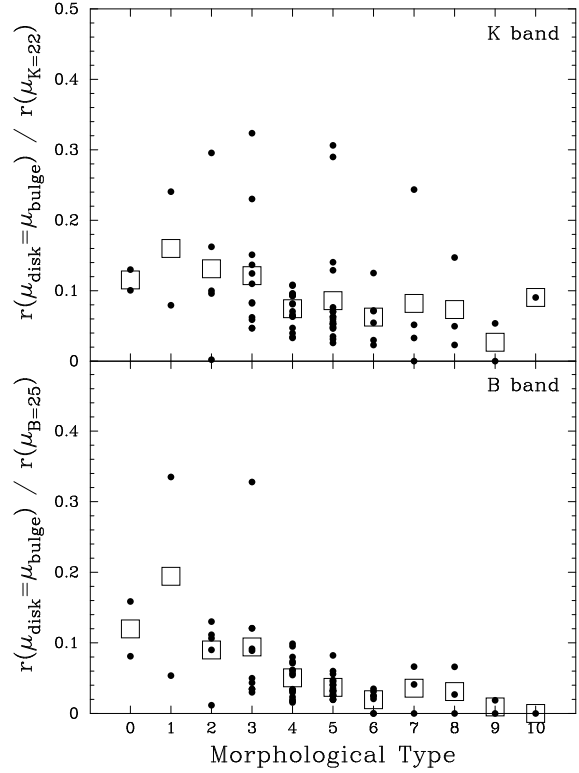,width=7.5cm,angle=00}}
\caption{The radius where the bulge and disk light contribute equally,
normalised by the indicated isophotal radius, is plotted against galaxy
morphological type.}
\label{SB2b-fig}
\end{figure}

\begin{figure}
\centerline{\psfig{figure=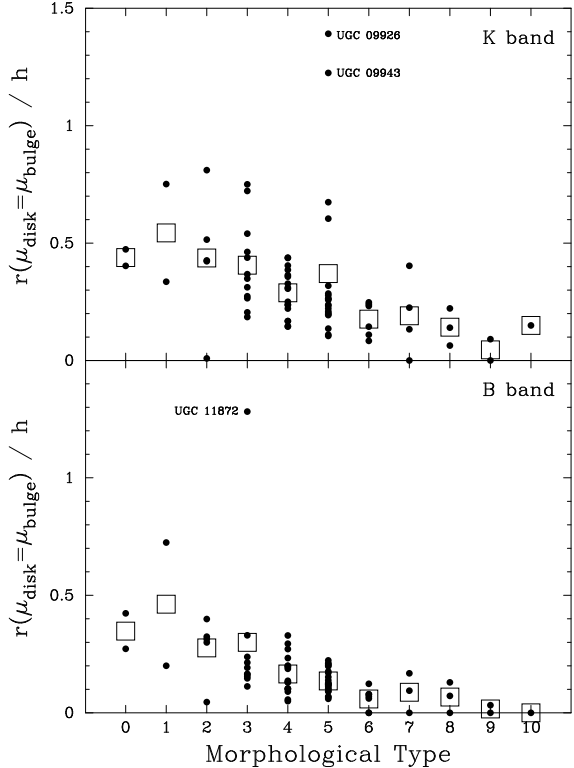,width=7.5cm,angle=00}}
\caption{The radius where the bulge and disk light contribute equally,
divided by the disk scale-length $h$, is plotted against galaxy
morphological type.}
\label{SB2c-fig}
\end{figure}

The second test uses the radius where the surface brightness of the disk and the 
bulge are equal; that is, the radius where the bulge starts to contribute less 
light than the disk.  To allow for different galaxy sizes, this radius was 
divided by the radius where the surface brightness of the
galaxy is equal to 22 $K$-mag for the $K$-band data, and 25 $B$-mag for 
the $B$-band data.  It was also separately normalised by dividing by the
scale-length of the disk. The results are presented in Figure~\ref{SB2b-fig}
and Figure~\ref{SB2c-fig}, and Table~\ref{stats-rad-h}.

%\placefigure{SB2b-fig}

%\placefigure{SB2c-fig}

% In Figure~\ref{SB2b-fig}, UGC 1559 ($T$$=$7) stands out.
% UGC 1559 is interesting because it is a late-type spiral whose bulge 
% clearly has a shape profile with $n$$<$1 but yet the central surface density
% of the bulge is significantly greater than that of the disk in the $K$-band 
% -- something which is unusual for late-type spirals. 

\placetable{stats-Mu}

\placetable{stats-rad-h}

%\placefigure{iceberg}
\begin{figure}
\centerline{\psfig{figure=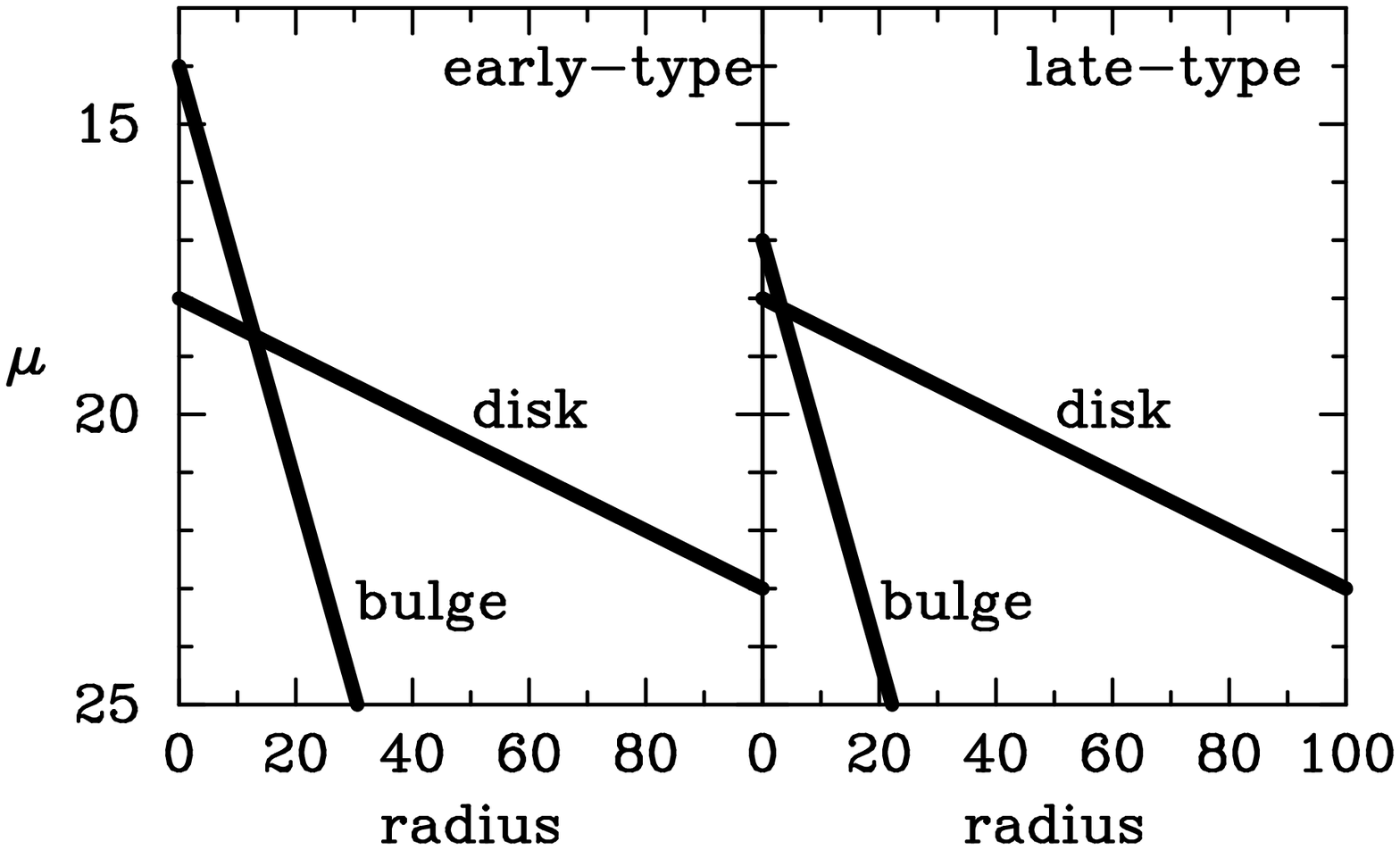,width=7.5cm,angle=00}}
\caption{This simple schematic illustrates how differences in the relative
stellar density between the bulge and disk can produce the optical illusion
that galaxies have different bulge-to-disk size ratios.  For the exponential models
used here, the slope is proportional to the scale-length; the bulge-to-disk
scale-length ratio is identical in both panels.}
\label{iceberg}
\end{figure}

The early-type galaxy sample used here does indeed `appear' to have, on average, 
larger bulges than the late-type sample.  So how does one interpret the $r_e/h$ data
which suggests that the relative size of the bulge and disk is independent of
morphological type?  The simple answer comes from a terrestrial analog -- icebergs.
If the bulge is somewhat submerged within the disk, achieved by turning down the 
bulge stellar density relative to the disk, then only the central peak of the 
bulge will be visible above the flux of the disk light. 
Increasing the bulge stellar density will rise the bulge up out of the disk, 
while the relative scale-length ratio remains unchanged.  This iceberg-like 
scenario is illustrated in Figure~\ref{iceberg}.  Of course, the difference
between bulges of early- and late-type spiral galaxies is more than a
case of adjusting the relative bulge/disk intensity, as the profile shapes 
also vary along the spiral Hubble sequence.  
However, as a rule,  the relative bulge/disk intensity appears to be a 
characteristic of the Hubble sequence of spiral galaxies.

% \section{Pitch Angle}

% \section{The contribution from Mercedes}

\section{Conclusions\label{SecCon}}

The shape parameters of the spiral galaxy bulges in this sample vary from 
$n$$\sim$0.5 to 4.  The results of previous studies which found such variation 
not to be random but to be systematic with Hubble type, such that late-type 
spiral galaxies have smaller shape parameters than early-type spiral galaxies, 
are confirmed.  Uncertainties on the model parameters, due to the uncertainty 
in the sky-background level, are small enough to exclude the possibility that 
all bulge profile shapes may be the same (i.e.\ for example $n$$=$4) and justify 
the need for using an $r^{1/n}$ bulge model.  Moreover, the $r^{1/n}$ bulge 
models result in a significantly better fit to the bulge than either the 
classical $r^{1/4}$ or exponential bulge model. This brings attention to the issue 
of simply using an $r^{1/4}$ law and exponential disk model for disky ellipticals, 
whose bulge is also likely to be described by an $r^{1/n}$ profile. 
Not only, in some cases not all, may the disk from such fits be an artifact, 
or left-over, from an inappropriately fitted $r^{1/4}$ law, but the bulge 
and disk model parameters will be in error to some degree from the forcing 
of a classical fitting function to the bulge. 

Fitting exponential light profile models to the bulges of spiral galaxies 
respectively under- and over-estimates the bulge sizes of early- 
and late-type spiral galaxies; to the degree that one obtains a 
smaller mean $r_e/h$ ratio for the former rather than the latter 
(Graham \& Prieto 1999a).  Fitting an $r^{1/n}$ bulge model reverse this 
result; however, not to the degree that the mean $r_e/h$ ratios are
significantly different between the early- and late-type samples. 
In contrast, the $B/D$ luminosity ratios are significantly 
($>$3$\sigma$) larger for the early-type spiral galaxies than the 
late-type spiral galaxies.  

It is the relative bulge-to-disk stellar density ratio, not size ratio,
that results in the apparent prominence of the bulge and trend with Hubble type. 
Although, for a given disk size, the disk surface brightness in late-type
spiral galaxies is fainter than in early-type spiral galaxies (Graham 2000), 
the surface brightness of the bulge in late-type spiral galaxies is yet 
fainter still, while the bulge-to-disk size ratio is the same.  One then 
has the picture for bulges in late-type galaxies as submerged beneath the 
surface brightness level of the disk -- somewhat akin to an iceberg scenario. 

A strong correlation exists between the shape of the bulge light profile 
and the $B/D$ luminosity ratio.  Spiral galaxy bulges do not all have 
exponential light distributions, less than 10\% do.  Similarly, the
correlation between the shape of the bulge light profile and the bulge
luminosity, in both the $B$- and $K$-bands, suggests that it may be the
mass of the bulge which dictates the stellar distribution in the bulge. 

That the Fundamental Plane (Djorgovski \& Davis 1987; Dressler et al.\ 1987) 
of elliptical galaxies and spiral galaxy bulges (Bender, Burstein, \& Faber 1992) 
are similar, despite their fundamental structural differences 
(Figure~\ref{fig-todo}), is intriguing.
The formation mechanisms at work in the Universe must be such that their structural
differences are compensated by dynamical differences in order to give rise to the 
physical scaling laws that define the Fundamental Plane.  This is very likely 
a reflection of the virial theorem (Faber et al.\ 1987), with the observed `tilt' 
explained by a combination of rotational support, non-homology in the velocity 
dispersion, and stellar population differences (Gregg 1992; Guzm\'{a}n \& Lucey 
1993; Bender, Saglia, \& Gerhard 1994; Prugniel \& Simien 1994, 1997; 
D'Onofrio, Longo, \& Capaccioli 1995; Pahre, Djorgovski, \& de Carvalho 1995; 
Ciotti, Lanzoni, \& Renzini 1996; Busarello et al.\ 1997; Graham \& Colless 1997; 
and a wealth of others).   
Why then, the dwarf elliptical galaxies, rather than the bulges of spiral 
galaxies, depart from the Fundamental Plane of ordinary elliptical galaxies 
adds to the intrigue, and may be explained when the above second order 
effects are fully dealt with for all classes of objects. 
Another question relating to the measured magnitude is the issue of 
extrapolation of the light profiles to infinity or the truncation at some 
isophotal radius possibly limited by the sky background light.

\acknowledgements
I wish to thank Mercedes Prieto for initiating this study, and 
Nicola Caon and Helmut Jerjen for providing me with 
the `ordinary' and dwarf elliptical galaxy data which was used 
to construct Figure~\ref{fig-todo}.  
This research has made use of the NASA/IPAC Extragalactic Database (NED)
which is operated by the Jet Propulsion Laboratory, California Institute
of Technology, under contract with the National Aeronautics and Space 
Administration.

\appendix
\section{$K$-band light profile data and models}

%\placefigure{figApp}
\begin{figure}
\centerline{\psfig{figure=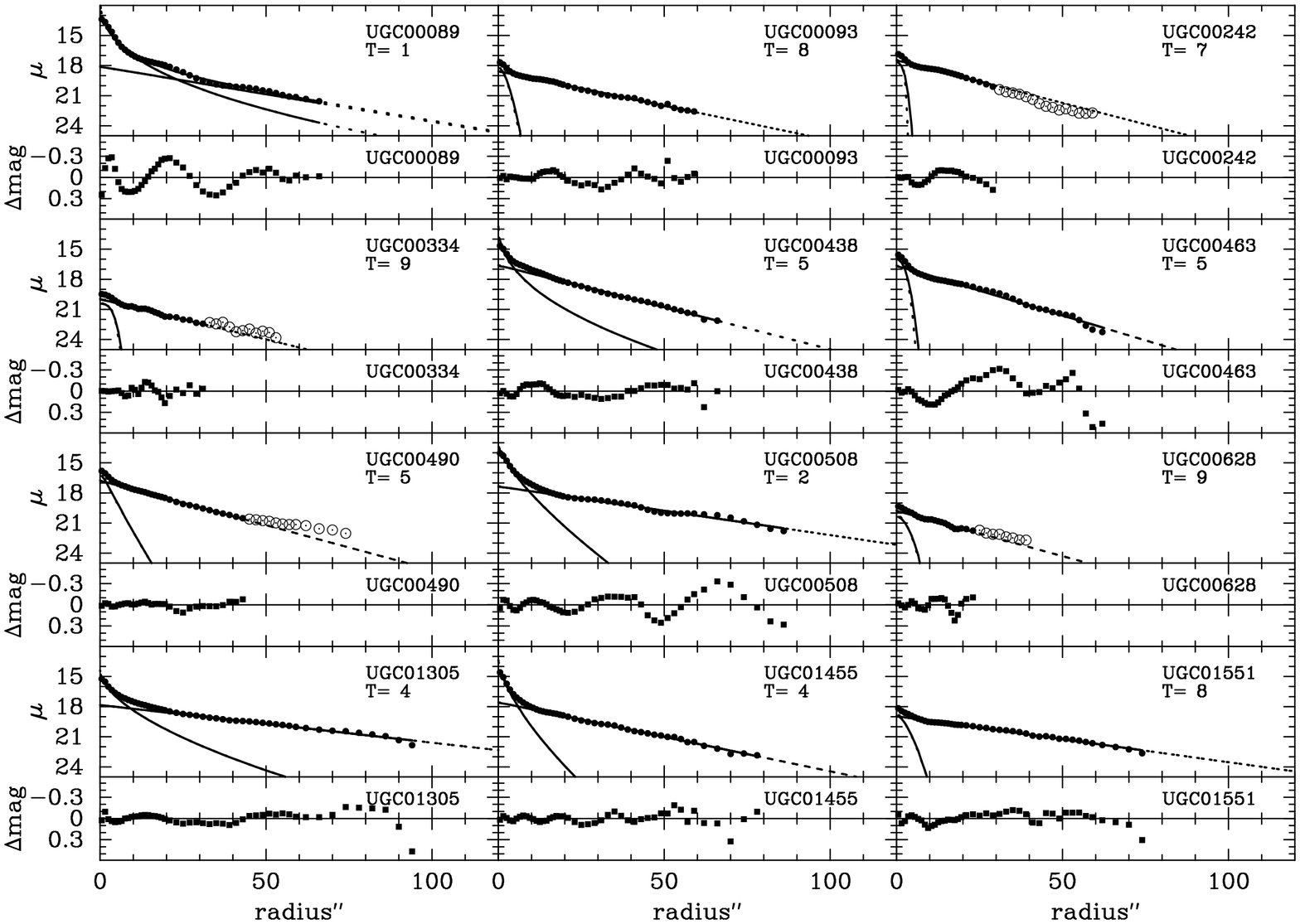,width=13.cm,angle=-90}}
\caption{The best-fitting seeing convolved $r^{1/n}$ bulge and exponential
disk models (solid lines) are fitted to the $K$-band surface brightness
profiles from de Jong (1996a).  The dashed lines are the models before
they are convolved with the PSF.}
\label{figApp}
\end{figure}

\begin{figure}
\centerline{\psfig{figure=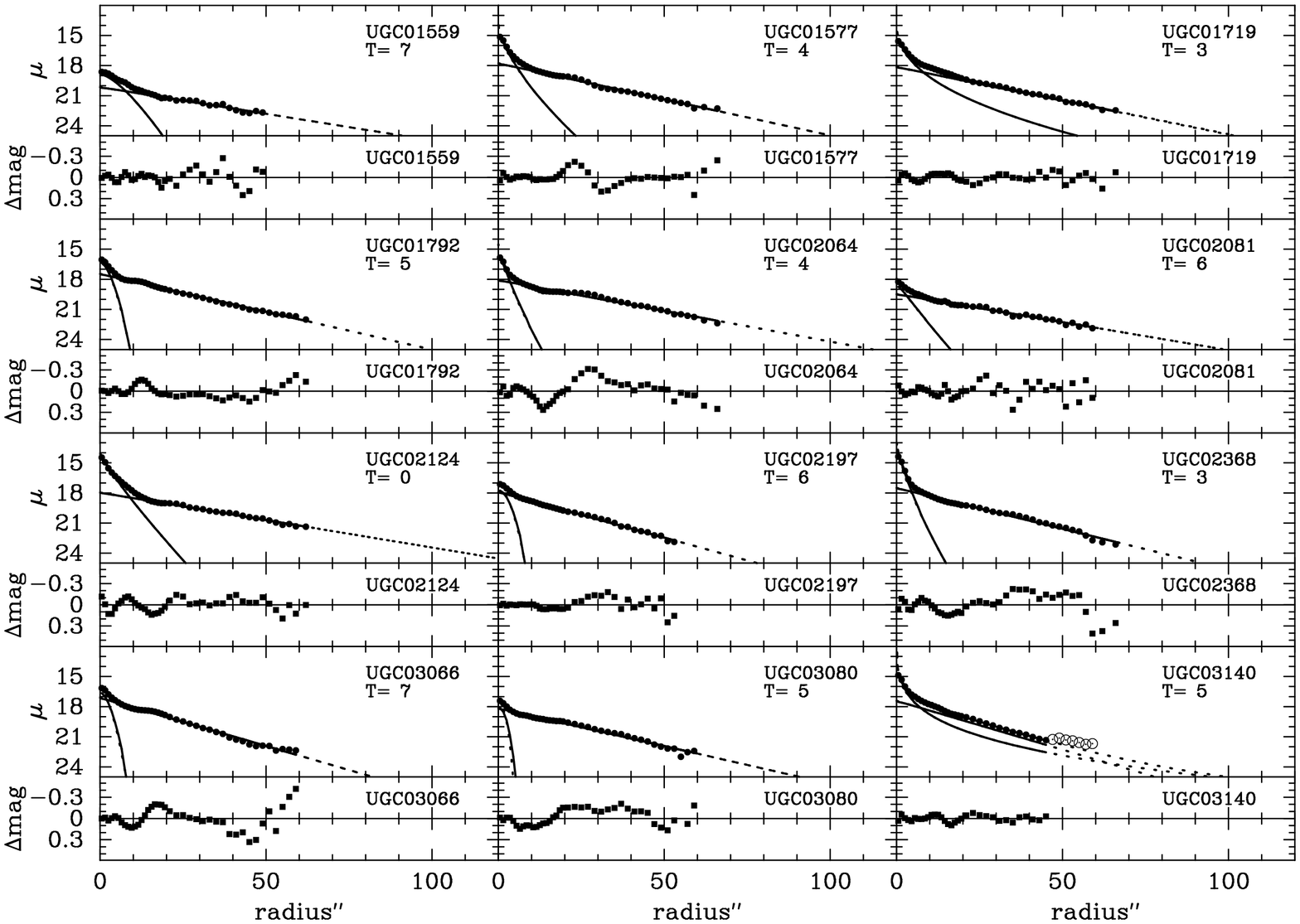,width=13.cm,angle=-90}}
\caption{{\it cont.}}
%\label{figApp_2}
\end{figure}

\begin{figure}
\centerline{\psfig{figure=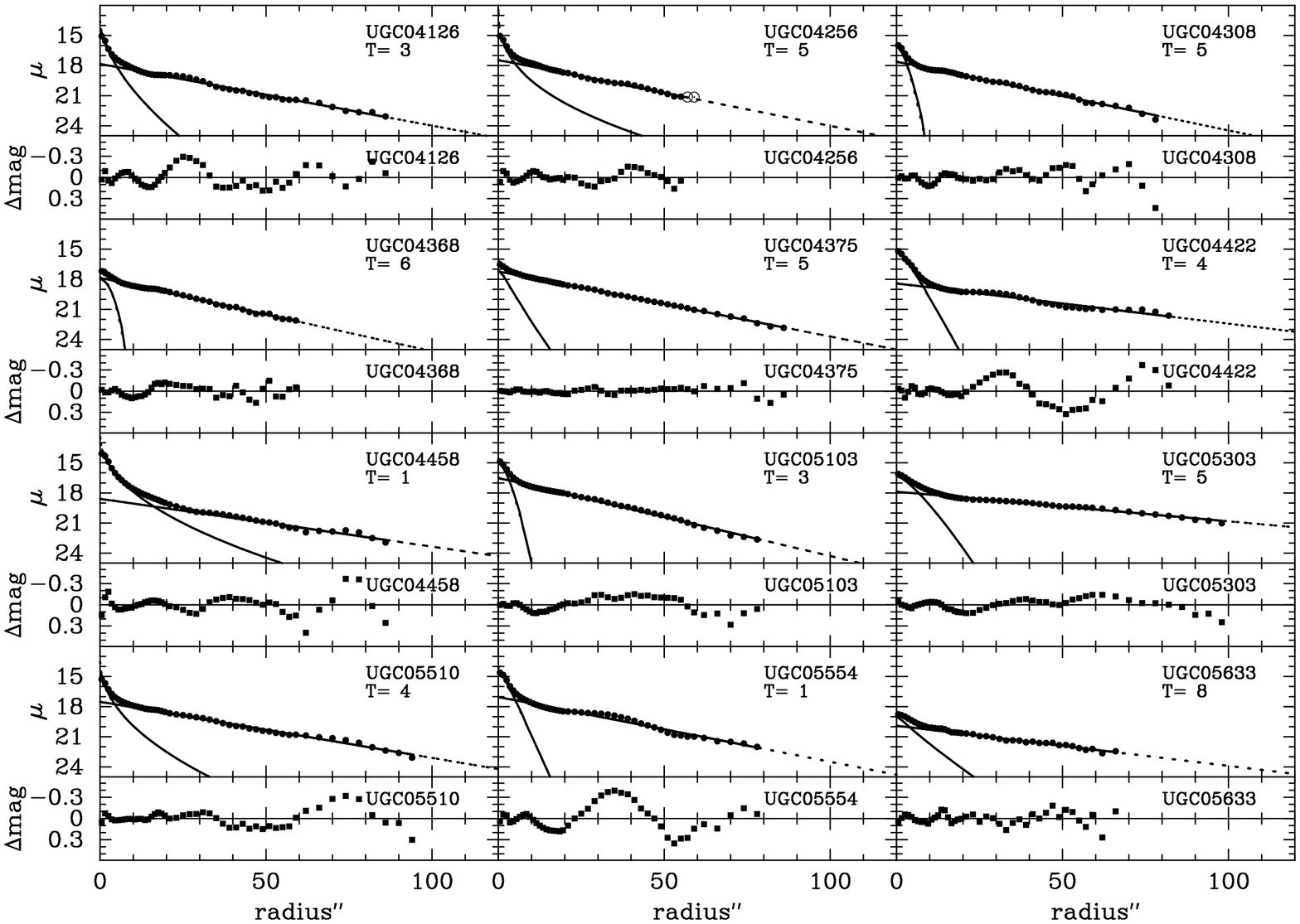,width=13.cm,angle=-90}}
\caption{{\it cont.}}
%\label{figApp_3}
\end{figure}

\begin{figure}
\centerline{\psfig{figure=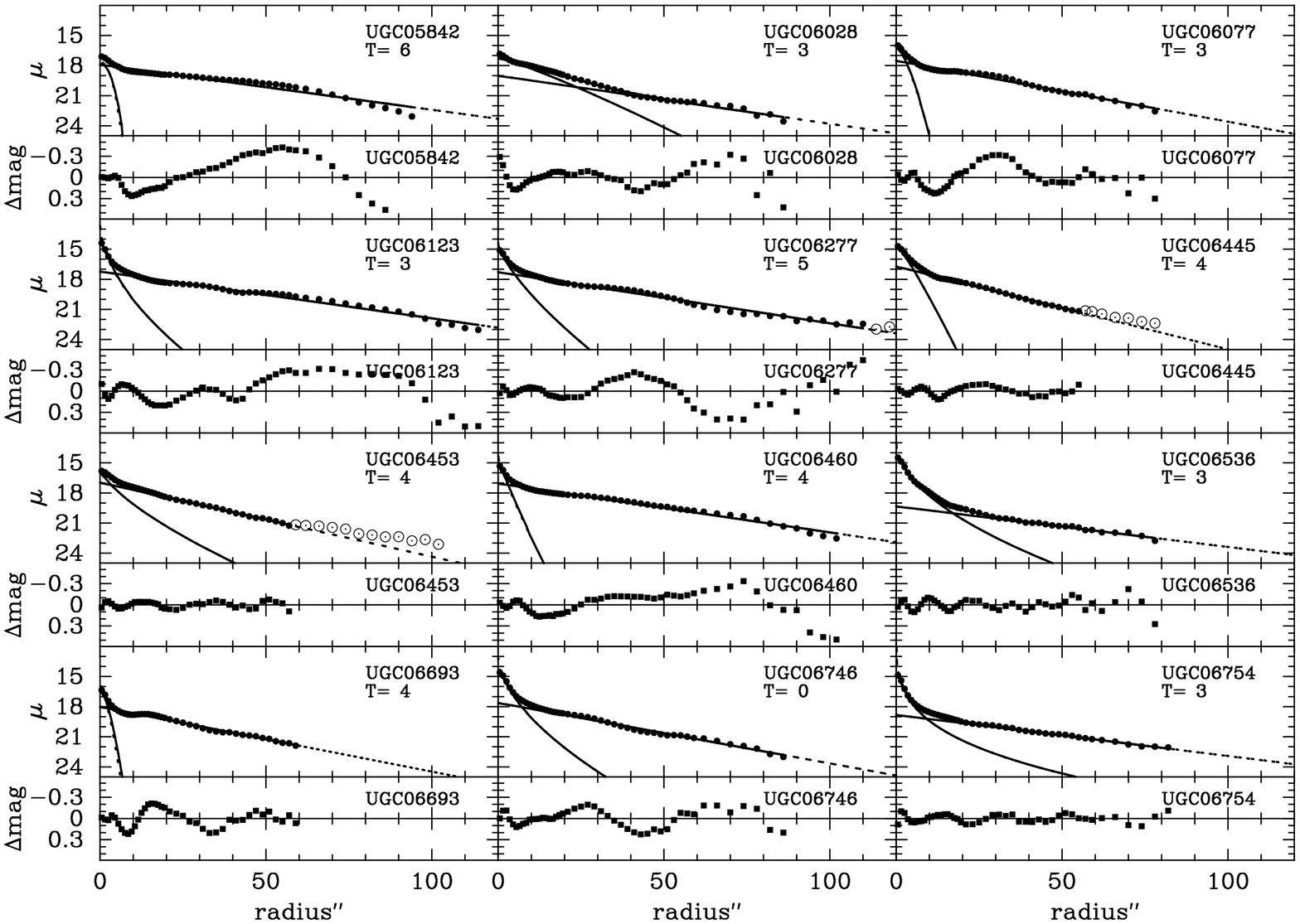,width=13.cm,angle=-90}}
\caption{{\it cont.}}
%\label{figApp_4}
\end{figure}

\begin{figure}
\centerline{\psfig{figure=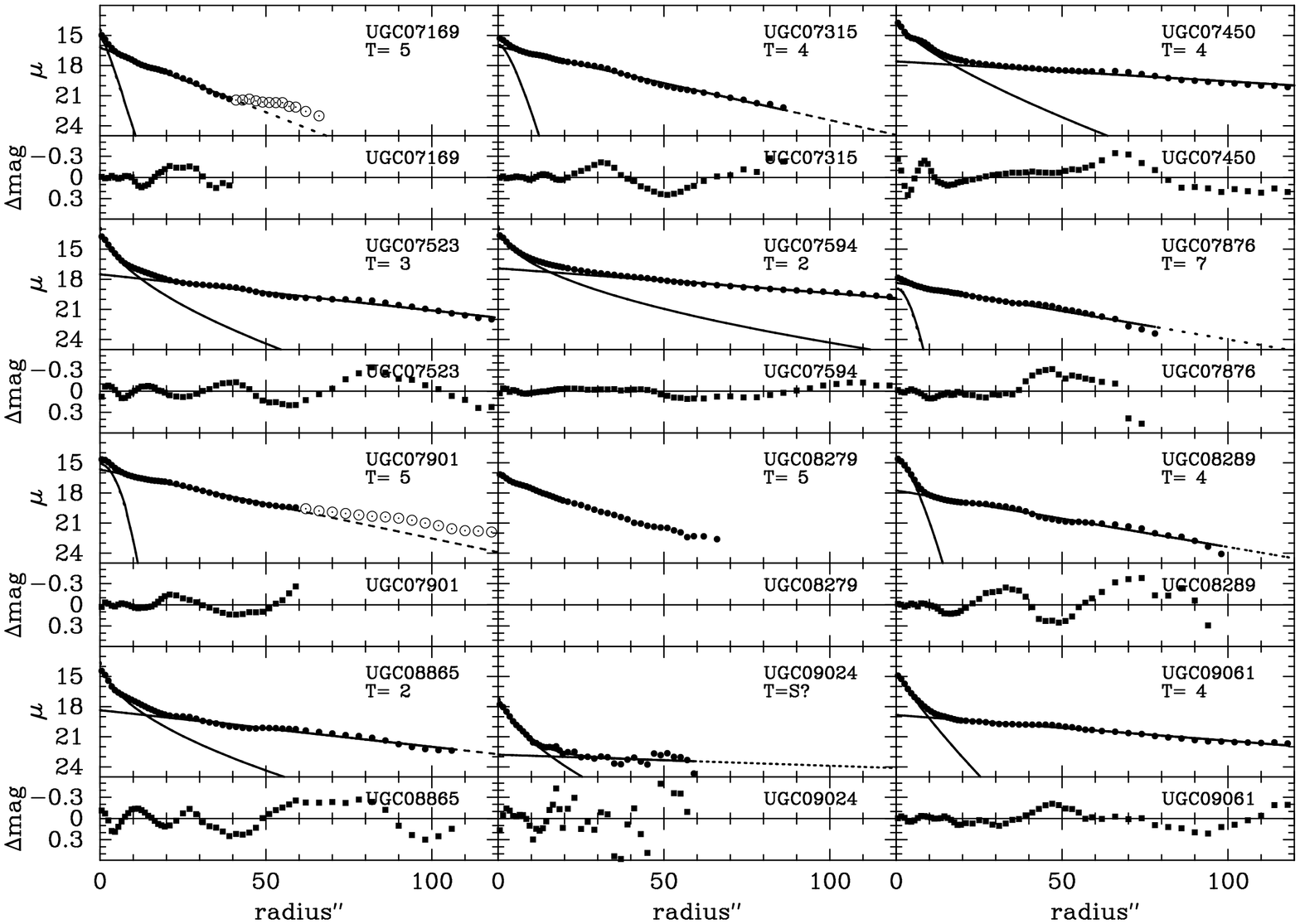,width=13.cm,angle=-90}}
\caption{{\it cont.}}
%\label{figApp_5}
\end{figure}

\begin{figure}
\centerline{\psfig{figure=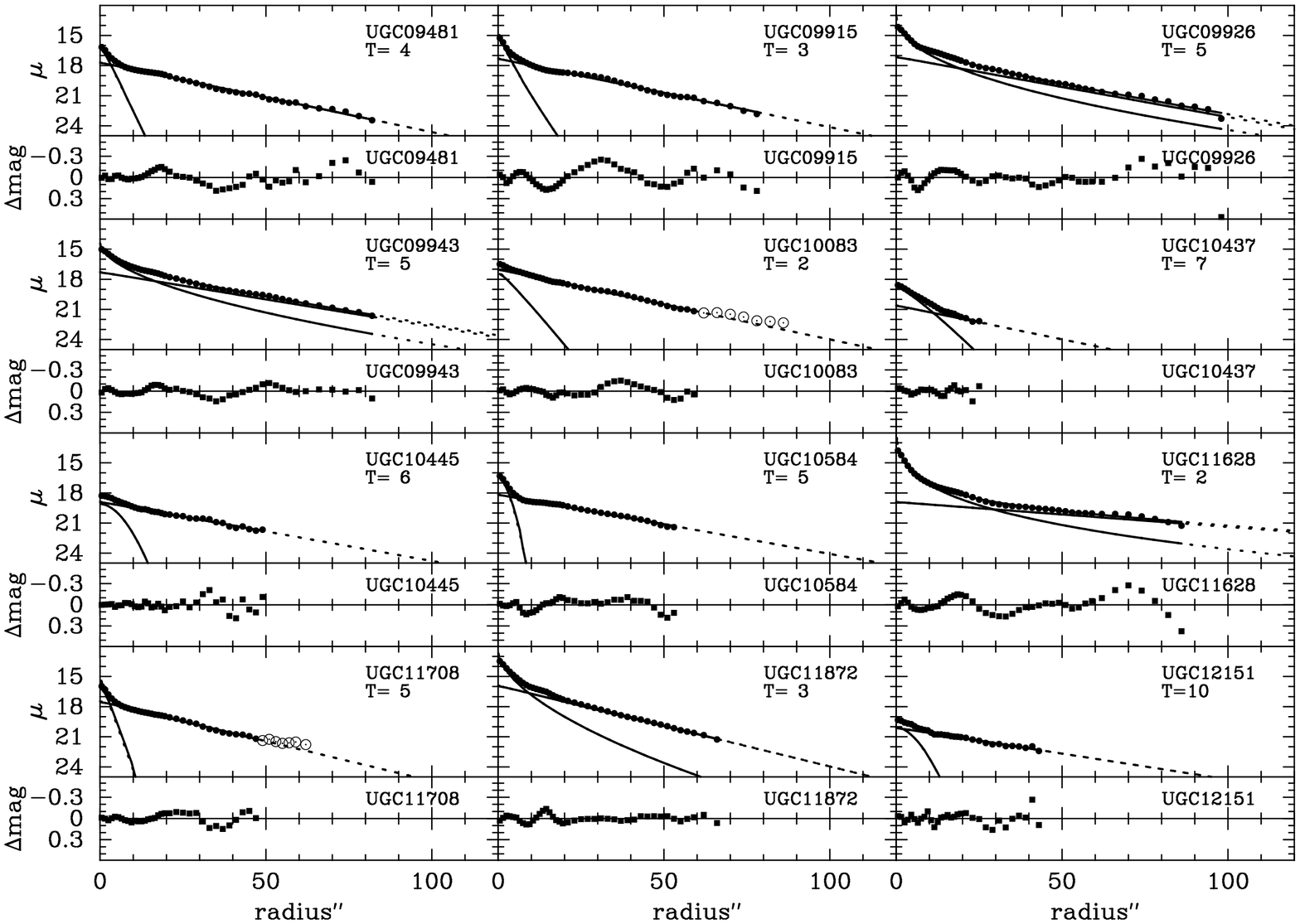,width=13.cm,angle=-90}}
\caption{{\it cont.}}
%\label{figApp_6}
\end{figure}

\begin{figure}
\centerline{\psfig{figure=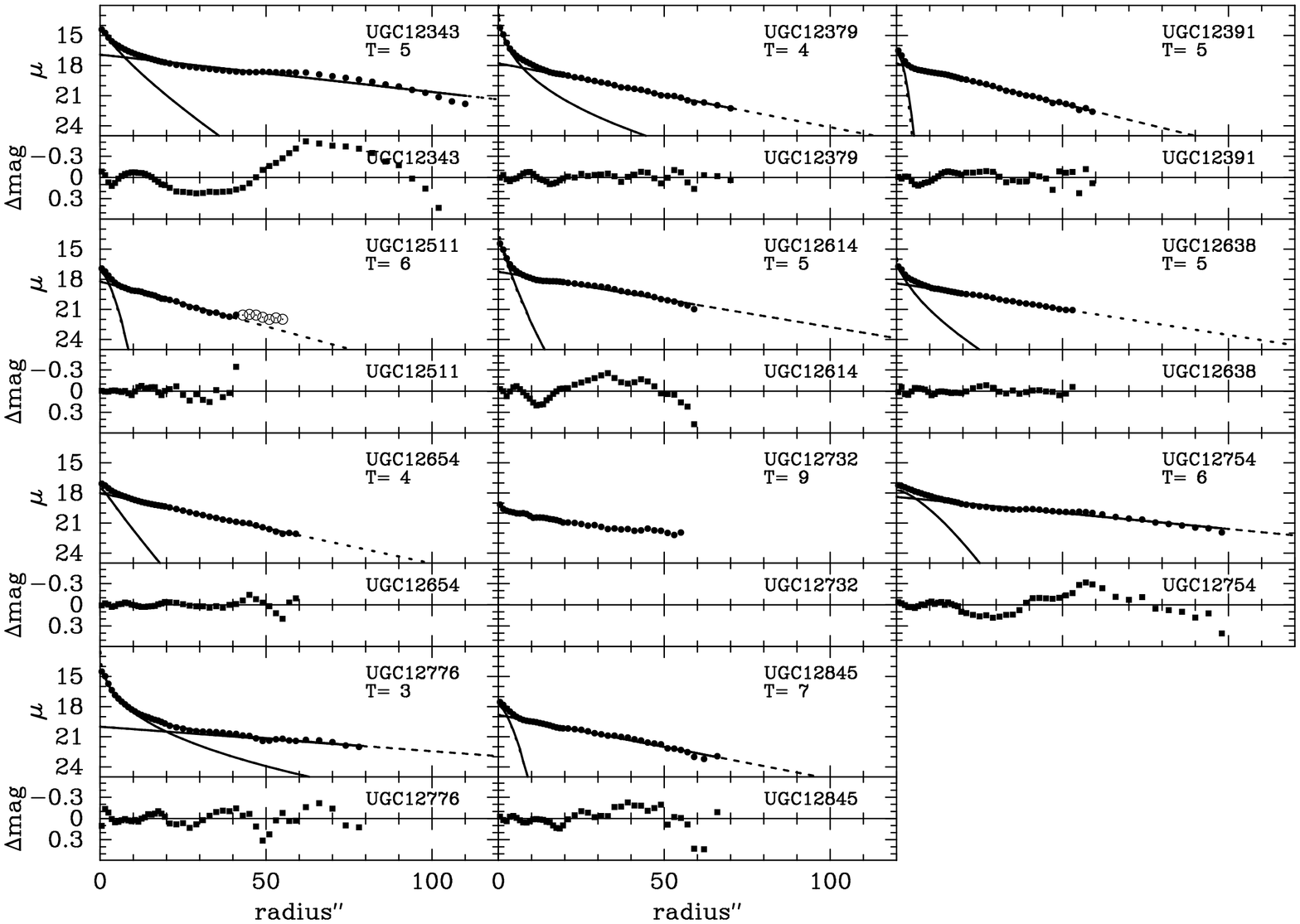,width=13.cm,angle=-90}}
\caption{{\it cont.}}
%\label{figApp_7}
\end{figure}

\newpage

\begin{deluxetable}{lcccc}
\tablewidth{0pt}
\tablecaption{Comparison of the $r_{e}/h$ data distributions\label{stats-r-h}}
\tablehead
{
\colhead{Band} &
\colhead{$\left<\frac{r_e}{h}\right>_{early-type} - \left<\frac{r_e}{h}\right>_{late-type}$} & \colhead{No.} & \colhead{Prob($F$)} & \colhead{Prob($t$), Prob($t_u$)}
}
\startdata
\multicolumn{5}{c}{Total Galaxy Sample} \\
B ..... & (0.211 - 0.130) = 0.081  & (20,16) & $<$.1\% & 08\%, 06\%  \\
R ..... & (0.209 - 0.149) = 0.060  & (20,19) & 01\%    & 06\%, 06\%  \\
I ..... & (0.241 - 0.183) = 0.058  & (15,18) & 01\%    & 13\%, 15\%  \\
K ..... & (0.213 - 0.199) = 0.014  & (19,18) & 89\%    & 58\%, 58\%  \\
 & & & \\
\multicolumn{5}{c}{Excluding galaxies modelled by de Jong to have a bar} \\ 
B ..... & (0.148 - 0.122) = 0.026  & (9,14) & 14\%   & 37\%, 42\%  \\
R ..... & (0.157 - 0.146) = 0.011  & (9,17) & 29\%   & 66\%, 62\%  \\
I ..... & (0.250 - 0.183) = 0.067  & (6,17) & 01\%   & 18\%, 37\%  \\
K ..... & (0.181 - 0.195) = --0.014 & (8,16) & 23\%  & 66\%, 61\%  \\
 & & & \\
\multicolumn{5}{c}{Excluding galaxies that had their outer profile truncated} \\
B ..... & (0.194 - 0.135) = 0.059  & (17,15) & $<$.1\%  & 19\%, 18\%  \\
R ..... & (0.207 - 0.167) = 0.040  & (18,15) &   0.5\%  & 27\%, 24\% \\
I ..... & (0.212 - 0.192) = 0.020  & (13,16) &    09\%  & 56\%, 56\% \\ 
K ..... & (0.214 - 0.203) = 0.011  & (18,14) & 65\%     & 75\%, 74\% \\
%  & & & \\
% \multicolumn{5}{c}{Excluding galaxies that which were not well modelled} \\
% B ..... & (0. - 0.) = 0.  & & \%  \\
% R ..... & (0. - 0.) = 0.  & & \%  \\
% I ..... & (0. - 0.) = 0.  & & \%  \\
% K ..... & (0.214 - 0.203) = 0.011 & & 39\%  38\%  \\  %T
\enddata
\tablecomments{Comparison of the $r_e/h$ data distributions for the early-type
(Sa,Sab,Sb) and late-type (Scd,Sd,Sdm,Sm) spiral galaxies. Column 1 shows the 
passband used.  The difference between the mean values of $r_e/h$ from the 
two distributions is shown in column 2, while Column 3 gives the number of 
galaxies in the early- and late-type samples respectively.  
Column 4 shows the probability, from an F-test, that the 
two distributions of $r_e/h$ have different variances (small values indicate 
significantly different variances -- in which case Students t-test with 
unequal variances should be used).  Column 5 gives the probability that 
the two distributions have the same mean value, as derived from Student's 
t-test assuming equal, Prob($t$), and unequal, Prob($t_u$), variances. Small 
probabilities indicate that the two data sets have significantly different 
means.}
\end{deluxetable}

\begin{deluxetable}{lcccc}
\tablewidth{0pt}
\tablecaption{Comparison of the $\log(B/D)$ data distributions\label{stats-B-D}}
\tablehead
{
\colhead{Band} &
\colhead{$\left<\log \frac{B}{D}\right>_{early-type} - \left<\log \frac{B}{D}\right>_{late-type}$} & \colhead{No.} & \colhead{Prob($F$)} & \colhead{Prob($t$)}
}
\startdata
\multicolumn{5}{c}{Total Galaxy Sample} \\
B ..... & ($-0.91 - -1.95$) = 1.04  & (20,16) & 31\% &  4.10$^{-06}$ \% \\ % , 2.10$^{-6}$ \%  \\
R ..... & ($-0.72 - -1.65$) = 0.93  & (20,19) & 88\% &  2.10$^{-05}$ \% \\ % , 2.10$^{-5}$ \%  \\
I ..... & ($-0.62 - -1.45$) = 0.83  & (15,18) & 40\% &  3.10$^{-04}$ \% \\ % , 6.10$^{-4}$ \%  \\
K ..... & ($-0.54 - -1.35$) = 0.81  & (19,18) & 91\% &  3.10$^{-04}$ \% \\ % , 3.10$^{-4}$ \%  \\
 & & & \\
\multicolumn{5}{c}{Excluding galaxies modelled by de Jong to have a bar}  \\ 
B ..... & ($-1.07 - -2.02$) = 0.95  & (9,14) & 19\% & 3.10$^{-03}$ \% \\ % , 4.10$^{-2}$ \%   \\
R ..... & ($-0.87 - -1.67$) = 0.80  & (9,17) & 97\% & 4.10$^{-02}$ \% \\ % , 7.10$^{-2}$ \%   \\ 
I ..... & ($-0.61 - -1.46$) = 0.85  & (6,17) & 20\% & 6.10$^{-02}$ \% \\ % , 1.10$^{-0}$ \%   \\ 
K ..... & ($-0.65 - -1.38$) = 0.73  & (8,16) & 62\% & 2.10$^{-01}$ \% \\ % , 6.10$^{-1}$ \%   \\
 & & & \\
\multicolumn{5}{c}{Excluding galaxies that had their outer profile truncated} \\
B ..... & ($-0.88 - -1.92$) = 1.04  & (17,15) & 23\% & 3.10$^{-05}$ \% \\ % , 3.10$^{-6}$ \%  \\
R ..... & ($-0.66 - -1.56$) = 0.90  & (18,15) & 83\% & 2.10$^{-04}$ \% \\ % , 2.10$^{-4}$ \%  \\
I ..... & ($-0.57 - -1.42$) = 0.85  & (13,16) & 61\% & 4.10$^{-04}$ \% \\ % , 8.10$^{-4}$ \%  \\ 
K ..... & ($-0.46 - -1.32$) = 0.86  & (18,14) & 64\% & 8.10$^{-04}$ \% \\ % , 2.10$^{-3}$ \%  \\ 
\enddata
\tablecomments{Comparison of the $\log(B/D$) data distributions for the early-
and late-type spiral galaxies.  Column 1 shows the passband used, while 
column 2 shows the difference between the mean $\log(B/D)$ values from the 
early- and late-type distributions.  Column 3 gives the number of galaxies 
in the early- and late-type samples.  Column 4 shows the probability that the 
two distributions have different variances (small values indicate 
significantly different variances).  Column 5 gives the probability that 
the two distributions have the same mean value -- as derived from Student's 
t-test.   The disk surface brightness has been corrected for 
inclination according to the prescripts given in the text.}
\end{deluxetable}

\begin{deluxetable}{lcccc}
\tablewidth{0pt}
\tablecaption{Comparison of the $\mu_{cen}-\mu_{\rm b=d}$ data distributions\label{Stats-Mu}}
\tablehead
{
\colhead{Band} &
\colhead{$\left<\log\frac{I_{\rm cen}}{I_{\rm b=d}}\right>_{early-type} - \left<\log\frac{I_{\rm cen}}{I_{\rm b=d}}\right>_{late-type}$} & \colhead{No.} & \colhead{Prob($F$)} & \colhead{Prob($t$), Prob($t_u$)}
}
\startdata
\multicolumn{5}{c}{Total Galaxy Sample} \\
B ..... & (-3.11 - -0.87) = -2.24 & (20,16) & $<$.1\% & 1.10$^{-5}$\%, 2.10$^{-5}$\% \\
R ..... & (-3.44 - -1.32) = -2.12 & (20,19) &   0.2\% & 5.10$^{-5}$\%, 1.10$^{-4}$\% \\
I ..... & (-3.64 - -1.43) = -2.22 & (15,18) & $<$.1\% & 4.10$^{-5}$\%, 2.10$^{-3}$\% \\
K ..... & (-3.84 - -1.48) = -2.36 & (19,18) & $<$.1\% & 2.10$^{-5}$\%, 1.10$^{-4}$\% \\
 & & & \\
\multicolumn{5}{c}{Excluding galaxies modelled by de Jong to have a bar} \\ 
B ..... & (-2.71 - -0.79) = -1.92 & (9,14)  & $<$.1\% & 9.10$^{-3}$\%, 4.10$^{-1}$\% \\
R ..... & (-3.03 - -1.32) = -1.72 & (9,17)  &   01\%  & 3.10$^{-2}$\%, 6.10$^{-1}$\% \\
I ..... & (-3.69 - -1.43) = -2.265 & (6,17)  & $<$.1\% & 5.10$^{-3}$\%, 2.10$^{-0}$\% \\
K ..... & (-3.69 - -1.43) = -2.26 & (8,16)  & 0.1\%   & 4.10$^{-3}$\%, 5.10$^{-1}$\% \\
 & & & \\
\multicolumn{5}{c}{Excluding galaxies that had their outer profile truncated} \\
B ..... & (-3.31 - -0.85) = -2.46 & (17,15) & $<$.1\% & 4.10$^{-6}$\%, 3.10$^{-5}$\% \\
R ..... & (-3.62 - -1.32) = -2.30 & (18,15) &    02\% & 8.10$^{-5}$\%, 6.10$^{-5}$\% \\
I ..... & (-3.86 - -1.44) = -2.42 & (13,16) &   0.1\% & 9.10$^{-6}$\%, 7.10$^{-4}$\% \\
K ..... & (-4.07 - -1.57) = -2.50 & (18,14) &   0.4\% & 3.10$^{-5}$\%, 2.10$^{-5}$\% \\
\enddata
\tablecomments{Comparison of the $(\mu_{\rm cen}-\mu_{\rm b=d})$ data distributions 
for the early-type (Sa,Sab,Sb) and late-type (Scd,Sd,Sdm,Sm) spiral galaxies. 
Column 1 shows the passband used.  The difference between the mean values of 
$\log(I_{\rm cen}/I_{\rm b=d})$ from the two distributions is shown in column 2, 
while Column 3 gives the number of galaxies in the early- and late-type samples. 
Column 4 shows the probability that the 
two distributions have different variances (small values indicate 
significantly different variances -- in which case Students t-test with 
unequal variances should be used).  Column 5 gives the probability that 
the two distributions have the same mean value, as derived from Student's 
t-test assuming equal, Prob($t$), and unequal, Prob($t_u$), variances. Small 
probabilities indicate that the two data sets have significantly different 
means.   The disk surface brightness has been corrected for 
inclination according to the prescripts given in the text.}
\end{deluxetable}

\begin{deluxetable}{lcccc}
\tablewidth{0pt}
\tablecaption{Comparison of the $r_{\rm b=d}/h$ data distributions
\label{stats-rad-h}}
\tablehead
{
\colhead{Band} &
\colhead{$\left<\frac{r_{\rm b=d}}{h}\right>_{early-type} - 
\left<\frac{r_{\rm b=d}}{h}\right>_{late-type}$} & \colhead{No.} 
& \colhead{Prob($F$)} & \colhead{Prob($t$), Prob($t_u$)}
}
\startdata
\multicolumn{5}{c}{Total Galaxy Sample} \\
B ..... & (0.312 - 0.061) = 0.251  & (20,16) & $<$.1\% & 3.10$^{-1}$\%, 3.10$^{-1}$\% \\
R ..... & (0.360 - 0.126) = 0.234  & (20,19) & $<$.1\% & 1.10$^{-2}$\%, 5.10$^{-2}$\% \\
I ..... & (0.404 - 0.164) = 0.240  & (15,18) & $<$.1\% & 5.10$^{-2}$\%, 1.10$^{-1}$\% \\
K ..... & (0.428 - 0.156) = 0.272  & (19,18) &    01\% & 9.10$^{-3}$\%, 5.10$^{-3}$\% \\
 & & & \\
\multicolumn{5}{c}{Excluding galaxies modelled by de Jong to have a bar} \\ 
B ..... & (0.268 - 0.050) = 0.218  & (9,14) & $<$.1\%  & 0.2\%, 3\%  \\
R ..... & (0.330 - 0.124) = 0.206  & (9,17) &   0.1\%  & 0.2\%, 4\%  \\
I ..... & (0.416 - 0.164) = 0.252  & (6,17) &   0.2\%  & 0.2\%, 5\%  \\
K ..... & (0.399 - 0.151) = 0.248  & (8,16) &    01\%  & 0.3\%, 2\%  \\
 & & & \\
\multicolumn{5}{c}{Excluding galaxies that had their outer profile truncated} \\
B ..... & (0.349 - 0.061) = 0.288  & (17,15) & $<$.1\% & 2.10$^{-1}$\%, 4.10$^{-1}$\% \\
R ..... & (0.389 - 0.137) = 0.252  & (18,15) & $<$.1\% & 3.10$^{-2}$\%, 4.10$^{-2}$\% \\
I ..... & (0.430 - 0.170) = 0.260  & (13,16) &   0.1\% & 5.10$^{-2}$\%  2.10$^{-1}$\% \\
K ..... & (0.463 - 0.170) = 0.293  & (18,14) &    05\% & 2.10$^{-2}$\%, 4.10$^{-3}$\% \\
\enddata
\tablecomments{Comparison of the $r_{\rm b=d}/h$ data distributions 
for the early-type (Sa,Sab,Sb) and late-type (Scd,Sd,Sdm,Sm) spiral galaxies. 
Column 1 shows the passband used.  The difference between the mean values of 
$r_{\rm b=d}/h$ from the two distributions is shown in column 2, 
while Column 3 gives the number of galaxies in the early- and late-type samples. 
Column 4 shows the probability that the 
two distributions have different variances (small values indicate 
significantly different variances -- in which case Students t-test with 
unequal variances should be used).  Column 5 gives the probability that 
the two distributions have the same mean value, as derived from Student's 
t-test assuming equal, Prob($t$), and unequal, Prob($t_u$), variances. Small 
probabilities indicate that the two data sets have significantly different 
means.  The disk surface brightness has been corrected for 
inclination according to the prescripts given in the text.} 
\end{deluxetable}

\end{document}